\documentclass[12pt]{article}
\usepackage{amsfonts,amsthm,amsmath,amssymb,upgreek,bm}
\usepackage[nosort]{cite}
\usepackage{hyperref}
\hypersetup{hidelinks}
\usepackage[paper=letterpaper,margin=.85in]{geometry}
\usepackage{graphicx}
\usepackage{units}
\usepackage{float}
\usepackage[export]{adjustbox}
\usepackage{xcolor}
\usepackage[shortlabels]{enumitem}
\usepackage[mathscr]{euscript}
\usepackage[normalem]{ulem}
\usepackage{bm}
\usepackage{physics}
\usepackage{tikz}
\usepackage{tikzsymbols}
\usepackage{pgfplots}
\usepackage{ytableau}
\usepackage{slashed}
\usepackage{mathtools}
\usetikzlibrary{decorations.pathmorphing}
\usetikzlibrary{arrows.meta}
\usepackage{comment}

\newcommand{\be}{\begin{equation}}
\newcommand{\ee}{\end{equation}}
\newcommand{\bea}{\begin{eqnarray}}
\newcommand{\eea}{\end{eqnarray}}

\renewcommand{\d}{\mathrm{d}}

\numberwithin{equation}{section}

\def\be{\begin{equation}}
\def\ee{\end{equation}}
\def\bea{\begin{eqnarray}}
\def\eea{\end{eqnarray}}
\def\ie{\begin{equation}\begin{aligned}}
\def\fe{\end{aligned}\end{equation}}

\newcommand{\cG}{{\mathcal G}}

\begin{document}
\thispagestyle{empty}

\vspace*{2.5cm}
\begin{center}

     {\LARGE \bf A non-prompt null ray cures its own singularity}

   \vspace{0.5in}
     
   {\bf  Gauri Batra}

    \vspace{0.5in}

   {Leinweber Institute for Theoretical Physics, Stanford University, Stanford, CA 94305, USA}  
                
    \vspace{0.5in}

    \vspace{0.5in}
    
\end{center}

\vspace{0.5in}

\begin{abstract}
    In AdS/CFT correlators, bulk cone singularities arise when two points that are timelike separated on the AdS boundary may be connected by a null geodesic through the bulk. We consider corrections to these correlators away from the $G_N\to 0$ limit arising from the backreaction of this null geodesic. A peculiar property of this geodesic is that it is not prompt, meaning that there exist points on it that are timelike separated. This implies that it may gravitationally affect itself. We show that this leads to an energy-dependent delay in the time elapsed along this geodesic.
    This coupled with possible inelastic effects due to gravitational radiation are the leading mechanisms that resolve the singularity. 
    The resolved singularity is accompanied by oscillations and hence zeroes in the real and imaginary parts of the correlator. We comment on the role that additional inelastic effects play in damping these oscillations.  Finally, we provide a concrete example in $2+1$D where we can compute the energy-dependent time delay effect explicitly by studying the boundary graviton mode.
\end{abstract}

\vspace{1in}

\pagebreak

\setcounter{tocdepth}{3}

\tableofcontents

\section{Introduction}
The AdS/CFT correspondence provides us with a way of experimentally probing quantum gravity. The main observables computed by this experiment are correlation functions of operators inserted on the AdS boundary. An interesting question is how these correlators encode information about quantum gravity. To make progress on this question, we can study the imprint of finite $G_N$ physics in the bulk on their analytic structure.

\begin{figure}
    \centering
    \includegraphics[width=0.3\linewidth]{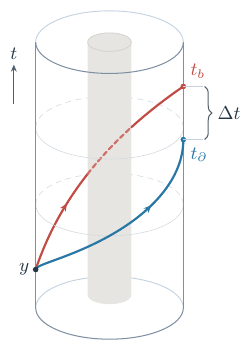}
    \caption{Bulk cone singularities of AdS/CFT correlators occur when two points that are timelike separated from the perspective of the boundary theory may be joined using a null geodesic through the bulk. This happens when the bulk has some matter content (pictured in grey). A null geodesic along the boundary that is emitted from the point $y$ is pictured in blue and arrives at the antipodal point at a time $t_{\partial}$. The null geodesic through the bulk (pictured in red) that starts at the same point $y$ arrives at the same antipodal point at a later time $t_b$. This leads to a singularity in a correlator where operators are inserted at the endpoints of the red geodesic.}
    \label{fig:bulk_cone_fig}
\end{figure}

In some cases, the gravity dual predicts that these correlators must be singular at points that do not lie on the light cone of the boundary theory. These singularities serve as laboratories to study how finite $G_N$ effects modify boundary observables.
A member of this collection of singularities is a class known as bulk cone singularities \cite{Hubeny:2006yu}.\footnote{A sampling of work on these kinds of singularities includes \cite{Dodelson:2023nnr,Leung:2024xsj,Dodelson:2025rng,Caron-Huot:2025she,Chen:2025cee,Caron-Huot:2025hmk,Haehl:2025ehf,He:2025yns,Auzzi:2025sep,Chen:2025ywj,Chen:2025jbf,Jia:2026pmv,Araya:2026shz,Jia:2026ryl}. Other classes of surprising singularities have also generated interest; see \cite{Fidkowski:2003nf,Festuccia:2005pi,Gary:2009ae,Maldacena:2015iua,Fitzpatrick:2016ive}.} These occur in asymptotically AdS geometries with matter. Their origin from the point of view of the gravity theory is not at all mysterious and arises from two basic facts: 
\begin{enumerate}
    \item The Gao-Wald theorem in AdS \cite{Gao:2000ga}, which considers null geodesics in geometries that are excited above vacuum AdS. In such a geometry, consider a null geodesic that travels through the bulk, connecting two points on the AdS boundary at times $t_i$ and $t_f$. The Gao-Wald theorem says that $t_f$ is greater than the time the null geodesic would have taken in empty AdS to arrive at the same spatial position on the boundary.
    \item On a fixed background, two-point functions of local fields are singular when their insertion points are connected by a null geodesic \cite{Landau:1959fi,Coleman:1965xm}.
\end{enumerate}
Let's consider the correlator $\ev{O(x)O(0)}_\Psi\equiv\ev{\Psi|O(x)O(0)|\Psi}$ as our observable, where $\Psi$ is an excited state of the boundary CFT and $x\equiv (t,\vec x)$. The second fact above implies that this correlator is singular when a null geodesic in the bulk joins the points $0$ and $x$.  From the point of view of the boundary, however, the first fact above says that these two points are timelike separated (see Figure \ref{fig:bulk_cone_fig} for an illustration). This implies that in the semiclassical limit when Einstein gravity governs the bulk, CFT correlators develop singularities off the light cone.

Previous work has studied the fate of such singularities at infinite $N$ and finite 't Hooft coupling $\lambda$ in the boundary theory. This regime corresponds to finite $\alpha'$ in the bulk. By considering a string following the trajectory of the null geodesic causing the bulk cone singularity, the authors of \cite{Dodelson:2020lal} found that the singularity gets shifted to a complex point in $t$. As a function of real $t$, the singularity turns into a finite-sized bump.

This work focuses on modifications to such singularities at finite $N$ in the boundary theory, which corresponds to a non-zero $G_N$ in the bulk. For simplicity, we will focus on the correlator at fixed spatial momentum $\vec p=0$, so that it is purely a function of the times at which the operators are inserted. We will consider effects that correspond to the gravitational backreaction of the null geodesic giving rise to the bulk cone singularity. Our main finding is that this backreaction fully resolves the singularity. In fact, these backreaction effects will be the dominant corrections responsible for curing the singularity.\footnote{Note that the correlator still receives important corrections from finite $\alpha'$ effects in $3+1$ bulk dimensions and higher --- $\alpha'$ effects are generally more important since to keep $g_s$ small, we need (for instance in AdS$_5\times$S$_5$) $\alpha'>>G_N^{1/4}$. Finite $G_N$ effects are responsible for transforming the shift we see at finite $\alpha'$ to a full resolution.} 

\begin{figure}
    \centering
    \includegraphics[width=0.3\linewidth]{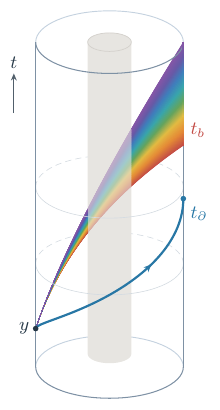}
    \caption{An illustration of one of the effects that resolves the bulk cone singularity at non-zero $G_N$ (or finite $N$ in the boundary CFT). When we go away from the $G_N\to 0$ limit, the backreaction of the null geodesic causing the singularity becomes important. This null geodesic experiences a time delay that depends on its energy; null geodesics of different energies form the rainbow shown in this picture. The effect is similar to the dispersion of light through a prism. The correlator that develops the singularity may be written as an integral over energies of null geodesics. Due to this energy-dependent time delay, high energies no longer add up in phase, leading to a finite answer.}
    \label{fig:rainbow}
\end{figure}

The basis of this resolution is twofold. First, the null geodesic under consideration experiences an additional time delay that is proportional to its energy. This time delay arises because this geodesic is not prompt, i.e., two points on it are timelike separated. As a result, part of this geodesic in the future is influenced by part of it in the past, and so experiences a change in its trajectory. This effect is visualised in Figure \ref{fig:rainbow}.\footnote{This is a mechanism more commonly known as gravitational self-force \cite{Mino:1996nk,Quinn:1996am} (for reviews see \cite{Poisson:2011nh}). It has been considered before in the AdS/CFT context in \cite{Shuryak:2011ge}.} Second, this null geodesic emits gravitational radiation with a certain probability. Projecting out final states with gravitational radiation in them also gives rise to a correction to the correlator at the same order.

From the point of view of the boundary conformal field theory, it is expected that these singularities are not present at finite $N$. In a unitary CFT$_2$ one can prove that at finite central charge, four-point functions are bounded away from the boundary light cone \cite{Maldacena:2015iua}. While writing this paper, we also learnt that Almheiri, Dodelson and Komargodski \cite{AlmheiriDodelsonKomargodski:2026} were able to place similar bounds on the correlator in higher-dimensional CFTs. The purpose of our work is instead to understand the physical mechanism behind the resolution of these singularities, and what that says about the functional form of the correlator in the vicinity of the resolved singularities.

A plan and short summary of this paper is as follows. In \textbf{Section \ref{sec:energy_integral}} we show how to write the correlator $\ev{O(t,\vec x)O(0)}_\Psi$ as an integral over energies of null geodesics. Using this integral representation we derive the presence of bulk cone singularities in the correlator for values of $t$ such that the times $0$ and $t$ on the boundary are joined by a $\vec p=0$ null geodesic through the bulk.

The singularity due to this null geodesic (christened $\eta$) arises from an integral of the form
\begin{align}
\label{eq:intro_integral_1}
    \ev{O(t) O(0)}_\Psi^{\vec p=0,\eta}\sim \int_0^\infty \d E\, E^{2\Delta-d} e^{-iET_b(\infty)}.
\end{align}
Here $T_b(\infty)$ is the time on the boundary at which the null geodesic must be emitted with momentum $(E,\vec p)$ so that it reaches the boundary again at time $t$. Importantly, $T_b(\infty)$ is independent of the energy of the null geodesic. Thus whenever points on the boundary at times $t$ and $0$ are joined by such a null geodesic through the bulk, $T_b(\infty)=0$ and the above integral diverges.

In \textbf{Section \ref{sec:resolution}} we give general arguments for why the backreaction of this null geodesic on the geometry resolves the singularity. As noted above, a null geodesic that travels through a spacetime with matter is not prompt, and parts of it are timelike separated. As a result, the trajectory of this null geodesic and hence $T_b(\infty)$ become energy-dependent. We show that in a region of $\mathcal{O}(\sqrt{G_N})$ around the singularity, this modifies the integral in \eqref{eq:intro_integral_1} to the form
\begin{align}
\label{eq:intro_integral_2}
    \ev{O(t) O(0)}_\Psi^{\vec p=0,\eta}\sim \int_0^\infty \d E\, E^{2\Delta-d} e^{-iET_b(\infty)+\frac{i}{2}c_\eta G_N E^2}.
\end{align}
Additionally, we show that inelastic effects due to gravitational radiation from the null geodesic generally give $c_\eta$ a positive imaginary part. 
This Gaussian correction renders the above integral convergent and resolves the singularity. Such a Gaussian correction has been known to resolve other kinds of singularities as well, such as the bulk point singularity \cite{Maldacena:2015iua}, forbidden singularities in AdS$_3$/CFT$_2$ \cite{Fitzpatrick:2016ive} and the bouncing geodesic singularity in two dimensions \cite{Batra:2026gbq} (see also \cite{Maldacena:2017axo} for a similar effect). 
We also find that this correction gives rise to prominent oscillations and hence zeroes in the real and imaginary parts of the correlator, which are subsequently damped due to further inelastic effects as we exit the $\mathcal{O}(\sqrt{G_N})$ region around the resolved singularity.

In \textbf{Section \ref{sec:toy}} we provide an example in $2+1$D where the energy-dependent time delay mechanism noted above can be seen explicitly. As a model of bulk matter we consider two point particles oscillating about the center of AdS$_3$. In this setup, tidal effects that correct the correlator at finite $\alpha'$ \cite{Dodelson:2020lal} are absent, and gravitational effects are the dominant ones.
We show how the bulk cone singularity appears in this model, and how the null geodesic of interest backreacts on the spacetime. In three spacetime dimensions, we can work in a gauge where gravitational effects are dictated by the boundary graviton mode. In this gauge, the null geodesic backreacts by causing a change in the embedding of the AdS boundary. We show how this leads to an energy-dependent change in the time separation of the boundary endpoints of this geodesic, changing the quantity $T_b(\infty)$ in a way that leads to the integral \eqref{eq:intro_integral_2}. We then also motivate why $c_\eta$ is expected to have a positive imaginary part.

We end in \textbf{Section \ref{sec:disc}} with a discussion of future directions.

\section{Bulk cone singularities from an integral over energies}
\label{sec:energy_integral}
In this section we will derive the bulk cone singularities by writing the correlator $\ev{O(x)O(y)}_\Psi$ as an integral over energies of null geodesics.\footnote{For a different derivation in the large mass approximation, see \cite{Hubeny:2006yu}. Such a large $E$ WKB approximation has also been used to derive the bouncing geodesic singularity in \cite{Afkhami-Jeddi:2025wra}.} Here $x$ and $y$ are points on the AdS boundary. The advantage of this method is that we do not need to take a limit where the dimension $\Delta$ of the operator $O$ is large. 

We will focus on asymptotically AdS geometries with a spherical boundary. We can first expand the dependence on $y$ in terms of Fourier modes in time and spherical harmonics on the sphere to write this correlator as:
\begin{align}
\label{eq:fourier}
    \ev{\Psi|O(x)O(y)|\Psi}
    = \int\frac{\d E}{2\pi}
    \sum_{\ell,a} e^{iEt_y} Y_{\ell a}^{*}(\Omega_y)\,
    C_{p}(x).
\end{align}
Here we defined $p\equiv(E,\ell,a)$, and wrote $y\equiv(t_y,\Omega_y)$.
$C_p(x)$ measures the overlap between a state created by the field insertion $O(x)$ and an eigenstate with fixed momentum $p$ along the AdS boundary. By using the path integral, it may be computed as
\begin{align}
\label{eq:probe_PI}
    C_{p}(x)
    \sim \frac{1}{Z}
    \int \d t_y\,\d\Omega_y\,Y_{\ell a}(\Omega_y)
    \int \frac{\mathcal{D}X\,\mathcal{D}\eta}{V_{\mathrm{diff}}}\,
    e^{i(S_{\mathrm{pp}}[X,\eta]-Et_y)},
\end{align}
where $S_{\rm{pp}}[X,\eta]$ is the action of the point particle $X$, and $\eta$ is the einbein field.\footnote{See Appendix \ref{sec:diagrams} for more details about this action.} 
In the rest of this section we will explain how to use this to compute $C_p(x)$ in terms of null geodesics. In various places we will use the tools introduced in \cite{Caron-Huot:2025hmk}. As a check, at intermediate steps we will apply our methods to the example of empty global AdS$_3$, where $C_p(x)$ is known exactly. 

\subsection{The WKB approximation in terms of null geodesics}
\label{sec:wkb}
Our analysis starts by making $x$ a general point in the bulk; to obtain the expression for the correlator we will eventually take $x$ to the boundary.
In the WKB approximation for large $E$, this path integral is given by \cite{Caron-Huot:2025hmk}
\begin{align}
\label{eq:wkb_Cp}
    C_p(x)\approx \sum_I A_I\sqrt{\mathcal{D}_I(p;x)} \exp(iS_I(p;x)).
\end{align}
Before justifying why we obtain this expression in the WKB approximation, let us unpack its constituents. The sum over $I$ represents the sum over saddle points, and the contribution of each saddle may be computed in terms of a null geodesic. Each null geodesic starts at the AdS boundary with momentum $p$ and reaches the point $x$; the endpoint on the AdS boundary may lie in the past or future of $x$. For example, in global AdS$_3$, some of the contributing geodesics are given by:
\begin{equation}
\label{eq:six_null_geodesics}
    \includegraphics[width=\linewidth]{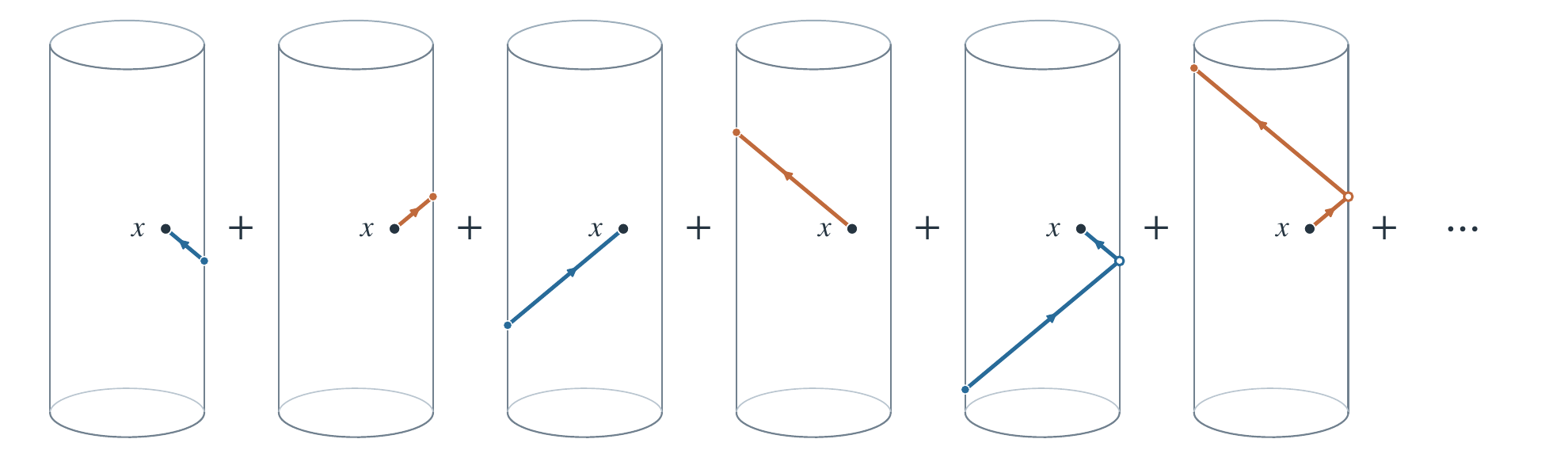}
\end{equation}
The full set of geodesics includes those that bounce multiple times off the AdS boundary before reaching their endpoint at the boundary.
In each case, only the endpoint $x$ of the null geodesic is fixed. $S_I$ is the on-shell action of this geodesic, given by
\begin{align}
\label{eq:action}
    S_I(p;x)=S_I (X_I,x)-ET_I+W_{\ell a,I}(\Omega_I).
\end{align}
Here $X_I\equiv(T_I,\Omega_I)$ is the second endpoint of the null geodesic that lies on the AdS boundary, which is computed by imposing that one endpoint is at $x$ and that the momentum of this geodesic at the second endpoint is $p$. The third term in the expression is the angular Hamilton–Jacobi generating function, which simplifies to $+L \phi_I$ in $2+1$D. For a null geodesic, the action $S_I (X_I,x)$ equals zero, so that the above expression simplifies to $S_I(p;x)=-ET_I+W_{\ell a,I}(\Omega_I)$.
The one loop factor $\mathcal D_{I}(p;x)$ is the van Vleck-Morette determinant evaluated on this geodesic (see Appendix \ref{sec:van_vleck} for more on this object), and $A_I$ is the normalization of each term. 

\subsubsection*{Why only null geodesics?}

To see why the WKB expression is given by that in \eqref{eq:wkb_Cp} and why we consider only null geodesics, we can work out the WKB analysis for a general metric with rotational and time translation symmetry:\footnote{A similar analysis may be found in \cite{Caron-Huot:2025hmk}.}
\begin{align}
    ds^2=-A(r) dT^2+\frac{dr^2}{B(r)}+r^2 d\phi^2.
\end{align}
For simplicity we have restricted to $2+1$ dimensions. Since we are computing the Wightman function, $C_p(x)$ satisfies the sourceless wave equation in momentum space:
\begin{align}
\label{eq:wave_eq}
    (\Box-m^2)C_p(x)=0.
\end{align}
Writing
\begin{align}
    C_p(T,r,\phi)=e^{-iET+i L\phi}\phi_{E,L}(r)
\end{align}
and plugging this into the wave equation for a massive scalar living on this metric gives
\begin{equation}
    \left[
        \frac{1}{r}\sqrt{B\over A}\,
        \partial_r\!\left(
            r\sqrt{AB}\,\partial_r
        \right)
        -\frac{L^2}{r^2}
        +{E^2\over A}
        -m^2
    \right]\phi_{E,L}(r)=0 .
\end{equation}
The two WKB solutions to this equation are then
\begin{align}
\nonumber
    \phi_{E,L}^{\pm}(r)
    &\simeq
    {\cal N}_{\Delta,E,L}^{\pm}
    \left(
        {1 \over r\sqrt{A(r)B(r)}\,P_r(r)}
    \right)^{1/2}
    \exp\!\left[
        \pm i\int_r^\infty dr'\,P_r(r')
    \right],\\
    P_r(r)&=\sqrt{{E^2-A(r)L^2/r^2\over A(r)B(r)}}.
    \label{app:WKBSolution}
\end{align}
Here we take $E\to\infty$ at fixed $m$ and $L/E$, so the mass term can be neglected at leading order away from the AdS boundary and radial turning points.
Now, let us see why this matches the semiclassical expression in Eq. (\ref{eq:wkb_Cp}). Up to a conventional factor of \(2\), the prefactor in parentheses is the van Vleck-Morette determinant computed in \cite{Caron-Huot:2025hmk}. To see that the argument of the exponent in the solution for $C_p(T,r,\phi)$ evaluates to $p_\mu X_I^\mu=-E T_I + L\phi_I$, we use that for a null geodesic
\begin{align}
    \frac{dT}{dr}=\mp\frac{E}{A(r) B(r) P_r(r)}, \quad \frac{d\phi}{dr}=\mp\frac{L}{r^2 B(r) P_r(r)},
\end{align}
so that
\begin{align}
    T-T_I=\pm\int_r^\infty \d r'\,\frac{E}{A(r') B(r') P_r(r')}, \quad
    \phi-\phi_I=\pm\int_r^\infty\d r'\,\frac{L}{r'^2 B(r') P_r(r')}.
\end{align}
These equations imply that the argument of the exponent simplifies to $-E T_I + L\phi_I$ for both the WKB branches --- this is the result expected from \eqref{eq:action}. In a general asymptotically AdS metric without the above symmetries, this says that WKB solutions near the boundary correspond to null geodesics with their associated  momenta. Since this part of the classical solution we are considering is a null geodesic, the entire geodesic (including the part deep in the bulk) must be a null geodesic as well.

\subsubsection*{The breakdown of WKB and matching to the exact solution}

Now, the WKB approximation leading to \eqref{eq:wkb_Cp} breaks down when $x\equiv (r,u)$ is very close to the AdS boundary, so that $r\geq |p|$. To deal with this, following \cite{Caron-Huot:2025hmk}, we patch the WKB contribution of each geodesic with the exact solution to the wave equation \eqref{eq:wave_eq} near the boundary. As we will see, this will also determine the normalization factors $A_I$. For simplicity we will specialize to $\ell=0$ geodesics with zero angular momentum. Let us focus on a null geodesic $n_I$ contributing to \eqref{eq:wkb_Cp} that lies to the past of $x$. Imagine transporting $x$ backwards along its trajectory until it almost reaches the endpoint of this geodesic at the AdS boundary:
\begin{equation}
    \includegraphics[width=0.8in]{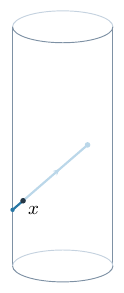}.
\end{equation}
In the overlap region where the geometry is asymptotically AdS and the WKB approximation is valid, the contribution of this null geodesic is given by
\begin{align}
\label{eq:near_boundary_wkb}
    C_{p,I}(r,u)
    \simeq \frac{A_I r^{-(d-1)/2}}{\sqrt{2E}}
    \exp\left[-iEt+\frac{iE}{r}\right],
\end{align}
where $u=(t,\Omega)$.
As we make $r$ larger so that the approximation above gets worse, this contribution transforms into a Hankel function that solves the asymptotically AdS wave equation \eqref{eq:wave_eq}:\footnote{Note that there exist other contributions that also transform into this Hankel function; here we focus on one of them.}
\begin{align}
    C_{p,I}(r,u)
    \simeq \frac{\sqrt{\pi}}{2} A_I
    e^{\frac{i\pi}{4}(2\Delta-d+1)}
    e^{-iEt}r^{-d/2}
    H^{(1)}_{\Delta-d/2}\left(\frac{E}{r}\right).
\end{align}
This can be checked by expanding the above solution in the overlap region, which reproduces 
\eqref{eq:near_boundary_wkb}.

Now, the above geodesic has a partner geodesic that has the opposite radial momentum and that lies to the future of $x$, pictured in red below:
\begin{equation}
    \includegraphics[width=0.8in]{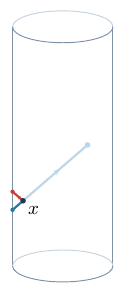}.
\end{equation}
The contribution of this partner geodesic instead behaves as
    \begin{align}
    \bar{C}_{p,I}(r,u)
    \simeq \frac{\sqrt{\pi}}{2}\bar{A}_I
    e^{-\frac{i\pi}{4}(2\Delta-d+1)}
    e^{-iEt}r^{-d/2}
    H^{(2)}_{\Delta-d/2}\left(\frac{E}{r}\right).
\end{align}
In the overlap region, the function above transforms into a null geodesic that moves towards the AdS boundary instead.
The contributions of these two geodesics must add up to give the form of $C_p(x)$ that leads to the universal short distance singularity of the $\vec p=0$ correlator:\footnote{To see why both contributions are needed, note that for the Wightman correlator, we focus on the normalizable solution. This is because it satisfies the wave equation with no boundary source \cite{Caron-Huot:2025hmk}. Both contributions are required to cancel the non-normalizable piece. This also fixes $
    \bar A_I\,e^{-\frac{i\pi}{4}(2\Delta-d+1)}
    =
    A_I\,e^{\frac{i\pi}{4}(2\Delta-d+1)}$.} 
\begin{equation}
   \lim_{r\to\infty} r^\Delta
    \left[C_{p,I}(r,u)+\bar{C}_{p,I}(r,u)\right]
    \sim e^{-iEt}
    \left(\frac{E}{2}\right)^{2\Delta-d}.
\end{equation}
In the above expression we omitted constant prefactors that are independent of $p$.
This condition then determines the $p$ dependence of the normalization prefactor $A_I$:
\begin{align}
\label{eq:normalization}
    A_I
    \sim
    \left(\frac{E}{2}\right)^{\Delta-d/2}.
\end{align}

\subsection{Singularities from the integral over null energies}
\label{sec:singularity_der}
To understand the singularities of the correlator using the technology derived in the previous subsection, we need to understand the quantity $C_p(x)$ of \eqref{eq:wkb_Cp} as we take $x$ to a general point on the AdS boundary. For simplicity we will work with the correlator at fixed spatial momentum $\vec p=0$, so that we only have to consider the subset of null geodesics  that satisfy this condition. This corresponds to the $\ell=0$ spherical harmonic. Let us focus on the contribution of a particular null geodesic $\eta$ to $C_p(x)$, where for now $x$ is a point close to the AdS boundary. This geodesic reaches $x$ from a point on the boundary in the past, with no intermediate reflections: 
\begin{equation}
    \includegraphics[width=0.8in]{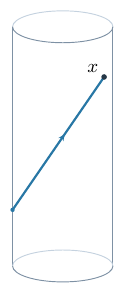}
\end{equation}
Consider its contribution to $C_p(x)$ in the overlap region where the geometry is asymptotically AdS and the WKB approximation is valid. It is given by
\begin{align}
\label{eq:eta_wkb}
    C_p^\eta(x)=A_\eta \sqrt{\mathcal{D}_\eta(p;x)}\exp(i S_\eta(p;x)), \quad S_\eta(p;x)=-E T_b(r,u).
\end{align}
Here we used \eqref{eq:action}, so that $T_b$ is the time on the boundary from which this null geodesic is emitted with energy $E$. Now, in the Hamilton-Jacobi formalism, we have that
\begin{align}
\label{eq:HJ_drS}
    \partial_r S\Big|_p=-E\partial_rT_b=P_r\sim \frac{k}{r^2}.
\end{align}
Here $k_\mu$ is the limiting tangential momentum of the null geodesic as we take $x$ to the AdS boundary\footnote{This is different from the momentum $p_\mu$ at the starting point of the geodesic, since the intermediate geometry may be time-dependent.} (and $k\equiv\sqrt{-g^{(0)}_{\mu\nu}k^\mu k^\nu}$, where $g^{(0)}_{\mu\nu}$ is the boundary metric), so that $P_r$ is determined by the null geodesic equation in the asymptotically AdS spacetime:
\begin{align}
    r^2 P_r^2-\frac{k^2}{r^2}=0.
\end{align}
Integrating \eqref{eq:HJ_drS}, we obtain
\begin{align}
    T_b(r,u)=T_b(r\to\infty,u)+\frac{k}{Er}.
\end{align}
Here $T_b(r\to\infty,u)$ (shortened to $T_b(\infty)$ from here on) is the starting point on the AdS boundary of the null geodesic that reaches the point on the boundary given by $(r\to\infty,u)$. Plugging this into \eqref{eq:eta_wkb} gives the following expression for the contribution of this null geodesic at large $r$:
\begin{align}
    C_p^\eta(x)=A_\eta\sqrt{\mathcal{D}_\eta(p;x)} \exp\left(-iET_b(\infty)-\frac{ik}{r}\right).
\end{align}
Writing it in this form makes it clear that we can again match this function to the required Hankel function. Doing so enables us to then take $x$ to the AdS boundary. To obtain a normalizable contribution, we then also need to add the contribution of the geodesic that has the opposite radial momentum, given by
\begin{equation}
    \includegraphics[width=0.8in]{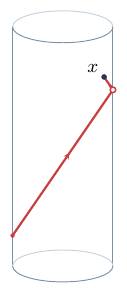}.
\end{equation}
Note that both these geodesics limit to the same geodesic as $r\to\infty$.
Adding the two and taking $x$ to the AdS boundary gives
\begin{equation}
\lim_{r\to\infty} r^\Delta
\left[
    C_p^\eta(r,u)
    + \bar C_{p}^\eta(r,u)
\right]\propto
E^{\Delta-\frac d2}\,
k^{\Delta-\frac{d-1}{2}}\,
e^{-iE T_b(\infty,u)}
\lim_{r\to\infty}
\left[
    r^{\frac{d-1}{2}}
    \sqrt{\mathcal D_\eta(p;r,u)}
\right].
\end{equation}
Here we omitted constant prefactors that do not depend on $E$, and used the normalization prefactor determined in \eqref{eq:normalization}. In Appendix \ref{sec:van_vleck}, we explain why $\sqrt{D_\eta}\propto E^{-1/2}$. Also using the fact that $k\propto E$,\footnote{This is because rescaling $E$ (which is the energy at the starting point of the null geodesic) amounts to changing the affine parameter along the null geodesic. Since the part of the null geodesic near $x$ (where it has momentum $k$) also satisfies the null geodesic equation, for $\vec p=0$ geodesics this amounts to rescaling $k$, implying that $k\propto E$.} the contribution of these terms to the 2-pt function at fixed $\vec p=0$ is given by
\begin{align}
\label{eq:bulk_cone_int}
    \ev{O(x) O(y)}_\Psi^{\vec p=0,\eta}\sim \int_0^\infty \d E\, E^{2\Delta-d} e^{-iET_b(\infty)+iEt_y}.
\end{align}
In the high energy region of this integral we can trust the WKB approximation that we have used. This region then gives rise to the divergence of the above 2-pt function as $t_y$ approaches $T_b(\infty)$, which is the required bulk cone singularity. Note that for this divergence, it was important that $T_b(\infty)$ is independent of $E$. 

\subsubsection*{A check using global AdS$_3$}

As a check, let us recover the singular behavior of the vacuum global AdS$_3$ correlator. Consider the analog of the geodesic $\eta$, which starts at the boundary, goes through the bulk, and ends at the boundary with no intermediate reflections. For this geodesic to end at a point $x\equiv(t,\phi)$ on the AdS boundary, it must start at $T_b(\infty)=t-\pi$. Thus its contribution to the fixed $L=0$ correlator is given by
\begin{align}
\label{eq:global_sing}
    \ev{O(t,\phi) O(0)}_\Psi^{L=0,\eta}\sim \int_0^\infty \d E\, E^{2\Delta-2} e^{-it E+i\pi E}\sim \frac{1}{(t-\pi)^{2\Delta-1}}.
\end{align}
This gives a singularity at $t=\pi$. To see why this is expected from the exact cylinder correlator, first note the exact expression 
\begin{align}
\label{eq:correlator_sum}
\ev{O(t,\phi) O(0)}_\Psi^{L=0}=
\sum_{n=0}^{\infty}
\left[
\frac{\Gamma(n+\Delta)}
{\Gamma(\Delta)\Gamma(n+1)}
\right]^2
e^{-i(\Delta+2n)(t-i\epsilon)}.
\end{align}
Now consider $t=m\pi+\delta$, with $\delta$ small. 
By using 
\begin{align}
    \frac{\Gamma(n+\Delta)}{\Gamma(n+1)}
\sim n^{\Delta-1},
\quad n\to\infty,
\end{align}
we can approximate the sum in \eqref{eq:correlator_sum} by the integral
\begin{align}
   \ev{O(m\pi+\delta,\phi) O(0)}_\Psi^{L=0}
\simeq
\frac{e^{-im\pi\Delta}}{\Gamma(\Delta)^2}
\int_0^\infty dn\,n^{2\Delta-2}
e^{-i(\Delta+2n)\delta}.
\end{align}
The large $n$ region of this integral evaluates to
\begin{align}
   \ev{O(m\pi+\delta,\phi) O(0)}_\Psi^{L=0}
\sim
\frac{1}
{
\delta^{2\Delta-1}},
\end{align}
giving singularities at $t=m\pi$, and matching the form of the singularity in \eqref{eq:global_sing} for $m=1$.
Other singularities of this correlator that correspond to $m> 1$ are reproduced by considering geodesics that bounce multiple times off the AdS boundary, such as in \eqref{eq:six_null_geodesics}.

\section{Resolution and properties of the resolved correlator}
\label{sec:resolution}
In the previous section, we showed why singularities appear in boundary CFT correlation functions when a null geodesic joins two points on the AdS boundary. In this section, we will outline the leading effects that contribute to resolving these singularities.
First, we will explain how the backreaction of this null geodesic on the geometry causes an energy-dependent change in its own trajectory. Then, we will outline the contribution of an inelastic effect due to the gravitational radiation from this null geodesic, which contributes at the same order. Together, these effects resolve the singularity.  We will then also study the properties of the correlator in the region surrounding the resolved singularity. The arguments in this section are general --- in Section \ref{sec:toy} we will show the backreaction effects studied in this section in a concrete example.

\subsection{The effects of backreaction}
\label{sec:backreaction}
Let us first summarize the basic idea of the previous section. For the fixed $\vec p=0$ correlator $ \ev{O(x) O(y)}_\Psi^{\vec p=0,\eta}$, we showed that the bulk cone singularity arises from the integral given in \eqref{eq:bulk_cone_int}. This integral is derived using the classical action of a null geodesic, and $T_b(\infty)$ is the time on the boundary at which this null geodesic is emitted so that it reaches the point $x$. 

When $G_N$ is non-zero, this null geodesic changes the geometry through which it propagates. This means that to compute the overlap that defines $C_p(x)$, the path integral in \eqref{eq:probe_PI} must be upgraded to
\begin{align}
\label{eq:Cp_grav}
    C_p(x)\sim \frac{1}{Z}\int\d t_y\,\d\Omega_y\,Y_{\ell a}(\Omega_y) \int \frac{Dg D\chi D X D\eta}{V_{\rm{diff}}} e^{i(S_{\rm{grav}}[g]+S_{\rm{matter}}[g,\chi]+S_{\rm{pp}}[X,\eta,g]-Et_y)}.
\end{align}
Here $S_{\rm{matter}}[g,\chi]$ is the action of the existing matter content in the geometry, with $\chi$ representing the matter field.
We will first describe how the action of the point particle $S_{\rm{pp}}^{(E)}[X,\eta]\equiv S_{\rm{pp}}-E t_y$ changes at non-zero $G_N$. An important property of a null geodesic propagating through the bulk in a matter state is that two points on it may be timelike separated.\footnote{In fact, this is a generic property of null geodesics in spacetimes with matter content \cite{Witten:2019qhl}. More precisely, assuming Einstein equations and the null energy condition, a complete null geodesic contains timelike-related points if $T_{\mu\nu}\dot X^\mu\dot X^\nu$ is nonzero somewhere along it.} When this happens, the geodesic is described as not being prompt (see \cite{Witten:2019qhl} for a clear exposition of this property). As a result, this geodesic gravitationally affects its own trajectory. It then experiences a time delay that is a function of its own energy.

With this new effect in mind, we can compute $C_p^\eta(x)$ using the same procedure as in Section \ref{sec:singularity_der}. That is, we can first write the corrected contribution of the geodesic $\eta$ in the WKB approximation, and then patch it to the exact solution near the boundary. Because of the backreaction, both $T_b$ and $\sqrt{D_\eta}$ in its contribution (given by the expression in \eqref{eq:eta_wkb}) receive corrections. Because of the time delay or advance noted above, $T_b$ receives a correction of the form $\Delta T_b(E)$, so that $S_{\rm{pp}}^{(E)}$ receives a correction of the form $-E\Delta T_b(E)$.

We then also need to upgrade the contribution of this null geodesic saddle point by including the changes in $S_{\rm{grav}}[g]$ and $S_{\rm{matter}}[g,\chi]$. These can be computed in a simple way in terms of the change in $S_{\rm{pp}}^{(E)}[X,\eta]$ as follows. Let $h_{\mu\nu}$ and $\delta\chi$ be the responses of the metric and matter fields to the null geodesic, and define
\begin{align}
    g_\lambda=g_0+\lambda h,\quad \chi_\lambda=\chi_0+\lambda\delta\chi
\end{align}
for some parameter $\lambda$.
Here $g_0$ and $\chi_0$ are the background values of the metric and matter fields, without the null geodesic. 
Since the background values satisfy the equations of motion of the gravity+matter action, we may write
\begin{align}
\label{eq:exp1}
    S_{\rm{bg}}\equiv S_{\rm{grav}}+S_{\rm{matter}}=S_{\rm{bg}}(0)+\lambda^2 \delta S_{\rm{bg}}+\cdots.
\end{align}
In the above expression we kept the $\mathcal{O}(G_N)$ correction to $S_{\rm{bg}}(0)$, using the fact that both $h_{\mu\nu}$ and $\delta\chi$ are small. To understand this more, let us work out how the second term above scales with $G_N$. To leading order in $G_N$, the change in the background action is of the form $\sim (h^2+h\delta\chi+(\delta\chi)^2)/G_N$.\footnote{Here we normalized the background matter fields in the action appropriately.} Since the stress tensor of the $L=0$ null geodesic is proportional to $E$, the linearized Einstein equations tell us that $h$ and $\delta\chi$ are both $\mathcal{O}(G_NE)$, so that the change in the action is $\delta S_{\rm{bg}}\sim G_N E^2$.

We also have
\begin{align}
\label{eq:exp2}
    S_{\rm{pp}}^{(E)}=S_{\rm{pp}}^{(E)}(0)+\lambda\delta S_{\rm{pp}}^{(E)}+\cdots.
\end{align}
To leading order, $\delta S_{\rm{pp}}^{(E)}$ is also $\mathcal{O}(G_N E^2)$, since $\Delta T_b(E)$ is $\mathcal{O}(G_NE)$ to leading order. To see why $\Delta T_b(E)$ scales in this way, let us start with the position space action of the massless point particle:
\begin{align}
    S_{\rm{pp}}[X,\eta,g]
    = \frac{1}{2}\int \d\lambda\,
    \frac{1}{\eta(\lambda)}\,g_{\mu\nu}(X)\dot X^\mu\dot X^\nu.
\end{align}
We can gauge fix the einbein field so that $\eta=1$.\footnote{The associated ghost fields decouple from $g_{\mu\nu}$ and $X^\mu$.} For a null path, this action evaluates to zero when evaluated on the equations of motion. This remains true after backreaction as well, so that $\delta S_{\rm{pp}}$ is also zero:
\begin{align}
    \delta S_{\rm{pp}}\sim \frac{1}{2}\int \d\lambda\, h_{\mu\nu}\dot X^\mu \dot X^\nu -\int \d\lambda\, (\mathrm{e.o.m.})\delta X^\mu+\dot X_\mu \delta X^\mu\Big|_i^f.
\end{align}
The last term above is evaluated on the initial and final endpoints of the null geodesic. Since this final endpoint is held fixed, the last term receives a contribution only from the initial endpoint. Evaluating it on the initial endpoint gives an answer proportional to $E\Delta T_b$. Using the fact that the second term is zero by the equations of motion, and that the first term is proportional to $G_NE^2$ then gives $\Delta T_b\sim G_N E$.\footnote{Here we also used that $d\lambda\sim E^{-1}$, and that $\dot X^\mu$ anywhere along the null geodesic is also proportional to $E$. This is because rescaling the initial energy $\dot X^\mu|_i\sim E$ amounts to rescaling the affine parameter, which rescales $\dot X^\mu$ anywhere along the null geodesic. }

Now, the full action $S_{\rm{total}}\equiv S_{\rm{bg}}+S_{\rm{pp}}^{(E)}$ must satisfy $\d S_{\rm{total}}/\d\lambda=0$ on setting $\lambda=1$. To leading order in $G_N$ where we keep only the $\mathcal{O}(\lambda^2)$ term in \eqref{eq:exp1} and the $\mathcal{O}(\lambda)$ term in \eqref{eq:exp2}, this gives rise to
\begin{align}
    \delta S_{\rm{bg}}=-\frac{1}{2} \delta S_{\rm{pp}}^{(E)}=\frac{1}{2} E \Delta T_b(E).
\end{align}
Taking these corrections into account, for non-zero $G_N$ the contribution of the geodesic $\eta$ in \eqref{eq:wkb_Cp} changes to
\begin{align}
\label{eq:corrected_Cp}
    C_p^\eta(x)=A_\eta \sqrt{\mathcal{D}_\eta^0(p;x)+\Delta \mathcal{D}_\eta(p;x)} \exp\left(-iE T_b(r,u)-\frac{1}{2} i E \Delta T_b(r,u)\right),
\end{align}
where we keep only the piece proportional to $G_N E$ in the expansion of $\Delta T_b(E)$. In Appendix \ref{sec:van_vleck} we also argue why in the $G_N$ expansion, $\Delta\mathcal{D}_\eta(p;x)\sim \sum_k G_N^k E^{k-1}$, implying that the terms in $\Delta\mathcal{D}_\eta(p;x)$ cannot sum up to a function that competes with the $G_NE^2$ dependence of the $\sim E \Delta T_b(E)$ correction at large $E$.\footnote{As we will argue at the end of this subsection, our main interest will be in $E\sim 1/\sqrt{G_N}$, so that this expansion is still valid.}

Putting these facts together and similarly carrying out the matching to the near boundary Hankel functions then gives the correction to $C_p^\eta(x)$ when $x$ lies on the AdS boundary. To leading order in $G_N$, the form of the corrected correlator becomes
\begin{align}
    \ev{O(x) O(y)}_\Psi^{\vec p=0,\eta}\sim 
\int_0^\infty dE\,E^{2\Delta-d}
\exp\left[
-iE\bigl(T_b(\infty;x)-t_y\bigr)
+\frac{i}{2}c_\eta G_NE^2
\right],
\end{align}
with $c_\eta$ real.\footnote{In an asymptotically AdS geometry, $c_\eta$ is also generically positive.} Note this is the form of the corrected correlator only in the vicinity of the would-be bulk cone singularity, since the contribution from the $\eta$ saddle point dominates in this region.
In the subsequent discussion, it is also simpler to study the above quantity by setting $y=0$ and $x=(t,0)$, which gives
\begin{align}
     \ev{O(t) O(0)}_\Psi^{\vec p=0,\eta} &\sim \int_0^\infty dE\,E^{2\Delta-d}\exp\!\left[-iE T_b(\infty;t)+\frac{i}{2}c_\eta G_NE^2\right].
     \label{eq:gaussian_smoothing}
\end{align}

\subsubsection*{Why can we focus on the Gaussian correction?}

Before we move on to the next section, let us explain why we only focused on the first $G_N$ correction to the exponent in \eqref{eq:gaussian_smoothing}. The reason is that doing so is well-controlled in the vicinity of the would-be singularity, where $T_b(\infty;t)\sim \sqrt{G_N}$ (which also corresponds to an $\mathcal{O}(\sqrt{G_N})$ region in $t$).
The starting point is to note that any corrections to the argument of the exponential in \eqref{eq:bulk_cone_int} (called $S_\eta$ from here) must organize into a series that has the following form at high energies:
\begin{align}
\label{eq:diagram_series}
    \Delta S_\eta\sim \sum_{k=1}^\infty G_N^k E^{k+1}.
\end{align}
Here we again set $\vec p=0$. To see this, we can compute the corrections to $C_p^\eta(x)$ using Feynman diagrams involving the null geodesic, the metric and the background matter.\footnote{We thank Simon Caron-Huot, Steve Shenker and Douglas Stanford for discussions related to such diagrams while \cite{Batra:2026gbq} was in progress, which helped with this analysis.}

We will expand around the solution where the null geodesic hasn't backreacted on the geometry. In Appendix \ref{sec:diagrams} we explain the coupling of the null geodesic to metric fluctuations in detail. In particular, we show that if we normalize the kinetic term of the gravitons so that propagators are $\mathcal{O}(1)$, a vertex involving a graviton anchored on the particle worldline gives rise to a factor of $\sqrt{G_N} E$. A vertex involving $p$ gravitons and the particle worldline instead gives rise to a factor of $G_N^{p/2} E$. Additionally, pure graviton vertices contribute more powers of $G_N$ --- a vertex with $n$ gravitons scales like $G_N^{n/2-1}$.\footnote{Here we assume the geodesic stays in regions of $\mathcal{O}(1)$ curvature. Since graviton vertices depend on the curvature and the background metric, a different treatment is required when the curvature is high and might scale with $E$.}

\begin{figure}
    \centering
    \includegraphics[width=\linewidth]{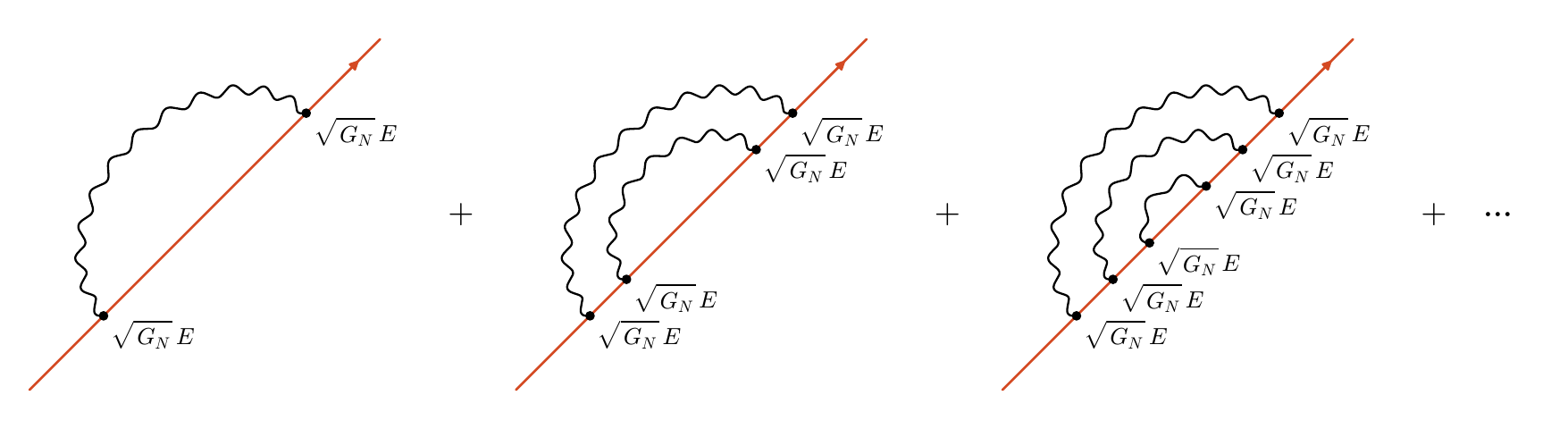}
    \caption{Some of the diagrams that exponentiate to give the correction to $S_\eta$ that is proportional to $G_N E^2$. Propagators are $\mathcal{O}(1)$, and the scaling of vertices is labelled. The straight orange line is the null geodesic, and the wavy lines are gravitons.}
    \label{fig:diagrams}
\end{figure}

Now, the correction $\Delta S_\eta$ is obtained from connected diagrams, whose disconnected repetitions exponentiate to give $\exp(\Delta S_\eta)$ (see Figure \ref{fig:diagrams}). This is the same combinatorial mechanism underlying eikonal exponentiation \cite{DiVecchia:2023frv}. Using the scaling of vertices and propagators above, a general connected diagram scales as $G_N^{k+v} E^{k+1}$, with $v\geq 0$.\footnote{Integrals over the null geodesic worldline cannot give additional factors of $E$, since the regulated null geodesic trajectory is independent of $E$. More precisely, if we parameterize the worldline by $t$, then the trajectory as well as the endpoints in $t$ are $E$-independent.} Thus the leading series at high energies and small $G_N$ has the form in \eqref{eq:diagram_series}.

To see why we may focus only on the first $G_N$ correction, first note that
in the region where $T_b(\infty;t)\sim \sqrt{G_N}$, the correlator is dominated by energies of $\mathcal{O}(1/\sqrt{G_N})$.\footnote{At larger energies $E\sim 1/G_N$, black hole formation effects make the integrand decay fast as a function of $E$. These effects may be modelled by imposing a cutoff on the integral at some $E\sim 1/G_N$.} In this region, the expansion \eqref{eq:diagram_series} is valid, and the term proportional to $G_N E^2$ dominates. This is why the correlator is given by the Gaussian integral in \eqref{eq:gaussian_smoothing}. On exiting this $\mathcal{O}(\sqrt{G_N})$ region in $t$, this description fails and additional terms in the full series become important. For instance, in an $\mathcal{O}(G_N^{1/4})$ region around $t_b$, we need to include the $\mathcal{O}(G_N^2)$ and $\mathcal{O}(G_N^3)$ terms in the series, and at an $\mathcal{O}(1)$ time away from $t_b$ the full series needs to be resummed. In the next section, we analyze some of these extra terms and comment on the real parts of their coefficients.

\subsection{The role of inelastic effects}
\label{sec:inelastic_effects}
In this section, we will argue how inelastic effects modify the analysis.\footnote{We thank Douglas Stanford for emphasizing the role of these effects and for helpful discussions that led to the content of this section. They have also been discussed in the context of OTOCs in \cite{Shenker:2014cwa}.} These effects reduce the probability amplitude of finding a single particle in the final state, and so multiply the $C_p^\eta(x)$ of \eqref{eq:corrected_Cp} by a factor that is less than one. This factor comes from imaginary parts in the diagrams considered in the previous subsection, which helps us read off how it scales with $G_N$ and $E$. The reason for this imaginary part is that cutting these diagrams gives rise to an on-shell process with a graviton in the final state. This means that there is a non-zero probability of a final state with a graviton in it, and hence the probability of having just a point particle in the final state is reduced.

\begin{figure}
    \centering
    \includegraphics[width=0.3\linewidth]{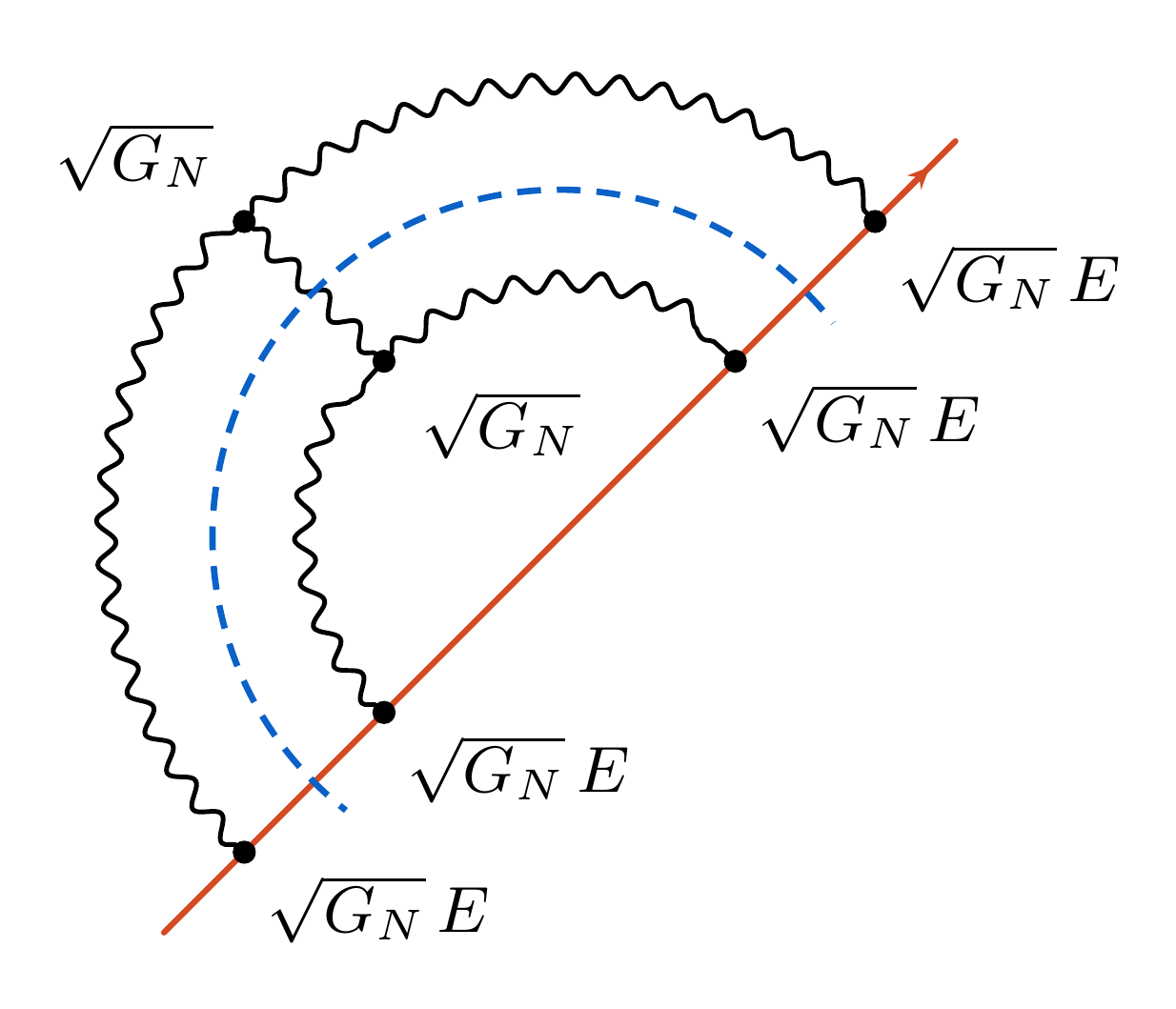}
    \caption{This figure pictures the generalization of the ``H diagram" of \cite{Amati:1987uf}. Instead of mediating the gravitational interactions between two particles, here the ``H" formed by the gravitons contributes to the gravitational interaction of a non-prompt null geodesic with itself. We have drawn the cut that gives rise to an intermediate on-shell graviton in blue. This inelastic effect corrects the correlator at $\mathcal{O}(G_N^3)$.}
    \label{fig:h_diagram}
\end{figure}

For the high energy scattering of two particles in flat space, the leading inelastic effect occurs at $\mathcal{O}(G_N^3)$ and is known as the ``H diagram" \cite{Amati:1987uf}. For null geodesics in a non-trivial background with matter, however, the leading inelastic effect is generally non-zero at $\mathcal{O}(G_N)$ because gravitational radiation from this null geodesic is generally an on-shell process.\footnote{Note that null geodesics in empty flat space or anti-de Sitter space don't radiate due to symmetry and kinematic constraints such as momentum conservation. In a general background with matter, interaction with the geometry can remove this obstruction (as an example, for the kick received by a black hole due to gravitational radiation see \cite{Favata:2004wz}).} An example of such an effect is gravitational radiation emitted from a massless particle orbiting a black hole. The amplitude for graviton emission (for a canonically normalized on-shell graviton $h_{\mu\nu}$) is given schematically by $\sqrt{G_N}\int \d^4x\, \sqrt{-g}\,h_{\mu\nu} T^{\mu\nu}$ (see \cite{Bernar:2017kug} for an example for a massive particle). Since $T_{\mu\nu}$ is proportional to $E$ (for a fixed null trajectory), the amplitude scales as $\sqrt{G_N}E$. As a result, the probability of emission scales as $G_NE^2$ for small $G_N$. Multiple emissions then exponentiate to give the probability $e^{-\# G_N E^2}$ of having no gravitons in the final state. A massless particle falling into a Schwarzschild black hole along a radial null geodesic also radiates \cite{Cardoso:2002ay}. 

This amplitude is exactly what we obtain if we cut the first diagram in Figure \ref{fig:diagrams} about its symmetry axis, which gives a diagram with a null worldline emitting a graviton. Hence if this gravitational radiation is allowed on-shell for the null geodesic under consideration, it gives an imaginary part to the diagrams in Figure \ref{fig:diagrams}. Imposing the lack of an on-shell graviton in the final state then multiplies $C_p(x)$ by  $e^{-\# G_N E^2}$, where the $G_N E^2$ scaling of the exponent is determined by the scaling of the first diagram in Figure \ref{fig:diagrams}. This inelastic effect thus gives a positive imaginary part to $c_\eta$. 

As an example, this effect would contribute to the resolution of bulk cone singularities that are related to the photon sphere \cite{Dodelson:2023nnr}, where the relevant null geodesic orbits a black hole in AdS. In Appendix \ref{sec:radiate} we also show explicitly that for an example null geodesic causing a bulk cone singularity in AdS$_5$-Schwarzschild, the amplitude to emit gravitational radiation is non-zero.
Note that we have not given a proof of the presence of this imaginary part for any null geodesic --- it would be interesting to understand the exceptions, if any.

We can also discuss the generalization of the H diagram to the case of a null geodesic in a non-trivial background. This is pictured in Figure \ref{fig:h_diagram}, along with the cut that generally gives rise to an on-shell graviton.  Summing up multiple insertions of the ``H" on the particle worldline then gives rise to a factor  of $e^{-\# G_N^3 E^4}$ multiplying $C_p(x)$, and so changes the integral in \eqref{eq:gaussian_smoothing} to
\begin{align}
\label{eq:inelastic}
    \ev{O(t) O(0)}_\Psi^{\vec p=0,\eta}\sim \int_0^\infty dE\,E^{2\Delta-d}\exp\!\left[-iE T_b(\infty;t)+\frac{i}{2}c_\eta G_NE^2+\gamma G_N^3 E^4\right],\quad \Re\gamma<0.
\end{align}
The important point is that the real part of the coefficient of the added $G_N^3E^4$ term is less than zero because of unitarity.

\subsection{Properties of the corrected correlator}
\label{sec:corrected_correlator}
We have argued that in the vicinity of the point where we expect to get a bulk cone singularity as $G_N\to 0$, the 2-pt function at fixed $\vec p=0$ is dominated by the integral
\begin{align}
\label{eq:more_gaussian}
     \ev{O(t) O(0)}_\Psi^{\vec p=0,\eta} \sim \int_0^\infty dE\,E^{2\Delta-d}\exp\!\left[-iE T_b(\infty;t)+\frac{i}{2}c_\eta G_NE^2\right],\quad \Im c_\eta\geq 0.
\end{align}
Here we renamed $T_b(\infty)$ to $T_b(\infty;t)$ to emphasize that it is a function of $t$.
Without any backreaction, the bulk cone singularity occurs at a value of $t$ ($\equiv t_b$) for which $T_b(\infty;t)=0$. This is because for $t=t_b$, all energies add up in phase. The additional Gaussian term coming from backreaction effects renders the integral convergent for any $t\to t_b$, resolving the singularity.\footnote{Note it is conditionally convergent when $\Im c_\eta=0$.}

Such a Gaussian smoothing has been known to resolve other kinds of singularities as well \cite{Maldacena:2015iua,Fitzpatrick:2016ive,Maldacena:2017axo}; it was studied in depth by the author in \cite{Batra:2026gbq} (see Section 4.2). In particular, the Gaussian smoothing of the singularity gives rise to two points surrounding the resolved singularity that locally look like the Airy function. Each of these points then acts as a source of three Stokes lines, that have interesting global behavior. Instead of repeating the saddle point analysis in \cite{Batra:2026gbq}, we will instead focus on aspects of the corrected correlator that are particularly interesting in this case. We will consider a situation in which the null geodesic is time delayed as a result of its backreaction, so that $\Re c_\eta>0$.

\begin{figure}[h]
    \centering
    \includegraphics[width=0.6\linewidth]{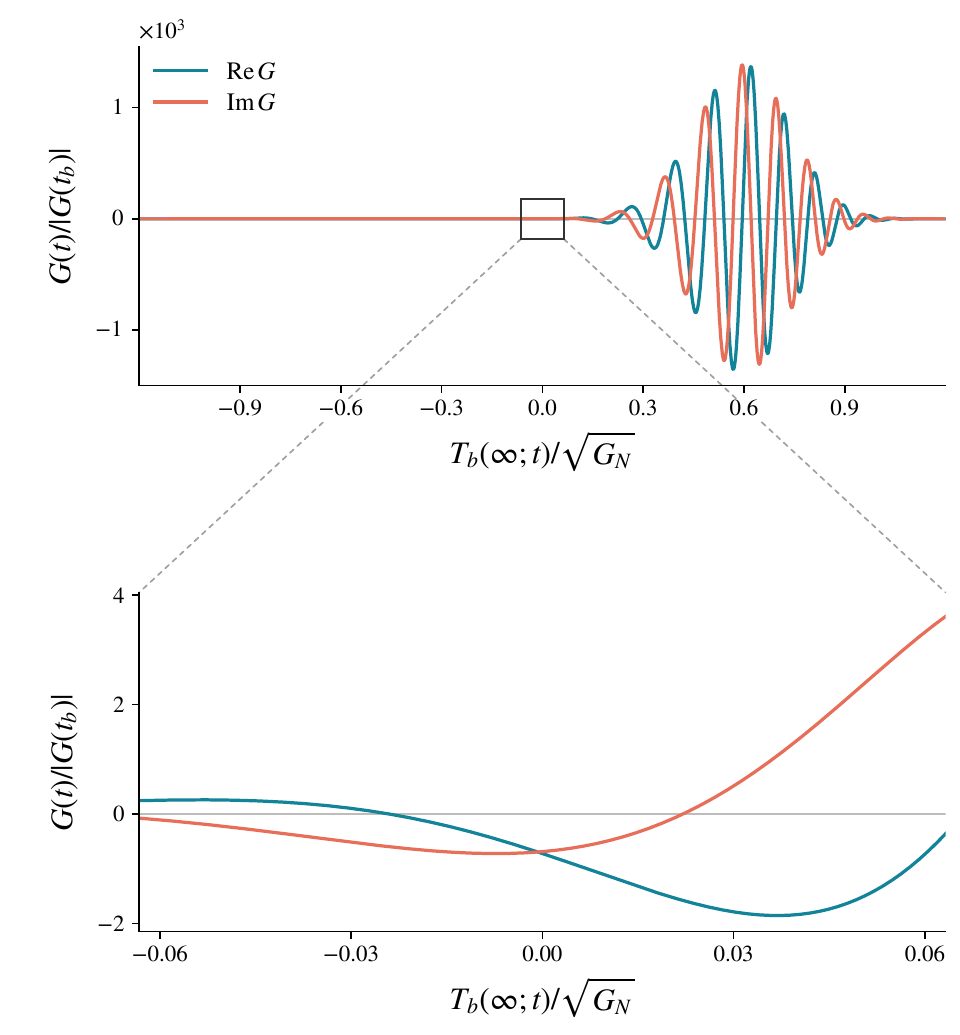}
    \caption{A plot of the correlator with backreaction of the null geodesic included, in an $\mathcal{O}(\sqrt{G_N})$ region around $T_b(\infty;t)=0$ (equivalently $t=t_b$, the location of the would-be singularity). Here we have normalized by the magnitude of the correlator at $t=t_b$. In this region, the correlator is given by the Gaussian integral in \eqref{eq:more_gaussian}, which may be done exactly (see \eqref{eq:parabolic}). Instead of resolving into a bump, the singularity resolves into a function that starts to oscillate after $t=t_b$. Here we defined $G(t)\equiv \ev{O(t) O(0)}_\Psi^{\vec p=0,\eta}$, and used $c_\eta=0.01+0.005 i$, $\Delta=9.7$ and $d=2$ as an example.}
    \label{fig:resolved_correlator}
\end{figure}

The integral in \eqref{eq:more_gaussian} is exactly doable and evaluates to
\begin{align}
\label{eq:parabolic}
\ev{O(t) O(0)}_\Psi^{\vec p=0,\eta}
\sim 
\frac{1}{
    (G_N)^{(2\Delta-d+1)/2}
}
\exp\left(-\frac{iT_b(\infty;t)^2}{4c_\eta G_N}\right)\,
D_{d-2\Delta-1}\!\left(
    e^{3i\pi/4}\frac{T_b(\infty;t)}{\sqrt{c_\eta G_N}}
\right),
\end{align}
where $D_\nu(z)$ is the parabolic cylinder function. Note that since this description is only valid in the region $T_b(\infty;t)\sim \sqrt{G_N}$, the above function scales like an inverse power of $G_N$. A plot of this function in the vicinity of $t=t_b$ is given in Figure \ref{fig:resolved_correlator}. We see that instead of resolving into a bump, the singularity instead gets resolved into a function that has pronounced oscillations (or zeroes) for $T_b(\infty;t)>0$ and hence $t>t_b$.\footnote{In the complex plane, the singularity cannot resolve into a bump around $t=t_b$ because of the maximum modulus principle.} These oscillations may generally be seen as arising from a saddle point of the exponent in \eqref{eq:more_gaussian}, which for $\Re c_\eta>0$ contributes to the integral over $E$ for $T_b(\infty;t)>0$ but not $T_b(\infty;t)<0$. It is located at
\begin{align}
    E_* = \frac{T_b(\infty;t)}{G_N c_\eta}.
\end{align}
The integral evaluated on this saddle point is given by \begin{align}
\ev{O(t) O(0)}_\Psi^{\vec p=0,\eta} \sim \sqrt{\frac{2\pi}{-iG_Nc_\eta}}\,
\left(\frac{T_b(\infty;t)}{G_Nc_\eta}\right)^{2\Delta-d}
\exp\!\left[-\frac{i\,T_b(\infty;t)^2}{2G_Nc_\eta}\right].
\end{align}
Note that the positive imaginary part of $c_\eta$ arising from inelastic effects damps these oscillations. As discussed in Section \ref{sec:inelastic_effects}, inelastic effects that contribute at higher orders in $G_N$ lead to terms in the integrand that are exponentially suppressed in $E$ as well. In Section \ref{sec:backreaction} we explained why these terms become important as we exit the region $T_b(\infty;t)\sim \mathcal{O}(\sqrt{G_N})$. The negative real parts of their coefficients then lead to a damping of the function as we exit the region $T_b(\infty;t)\sim \mathcal{O}(\sqrt{G_N})$ as well. 

An interesting property of the function in \eqref{eq:parabolic} is that for the physically relevant regime in which $\Im c_\eta>0$, it grows in a complex direction where $\Im t>0$ (or the direction of negative Euclidean time evolution). It would be interesting to see if the function keeps growing and becomes exponentially large in $1/G_N$ as we exit the $\mathcal{O}(\sqrt{G_N})$ region around $t=t_b$.\footnote{Such a behavior happens in the case considered in \cite{Batra:2026gbq}, where the exponentially large region of the function encodes information about the physics near the black hole singularity.} Since this exponential in $1/G_N$ growth must come from a saddle point of the $E$ integral at high energies, one would expect that this part of the function probes processes at high energies, such as black hole formation.

\subsubsection*{An optics analogy}
\begin{figure}
    \centering
    \includegraphics[width=0.8\linewidth]{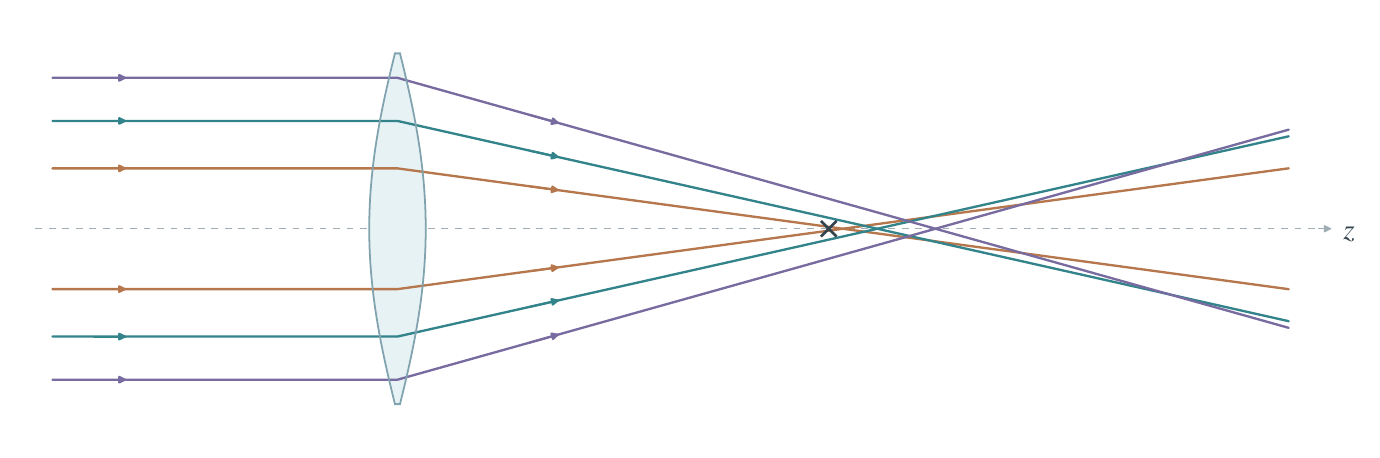}
    \caption{The setup in this paper may be mapped to a setup involving a lens with spherical aberration. For an ideal lens, light rays all focus onto a single point (marked with a cross). With spherical aberration, light rays hit the $z$ axis after the focus instead, mirroring the time delay experienced by a non-prompt null geodesic. Because of the resulting interference effects, the electric field amplitude (whose squared magnitude gives the intensity of light) on the $z$ axis close to the focal point is then given by an integral like \eqref{eq:more_gaussian}.}
    \label{fig:lens}
\end{figure}

The behavior of the resolved singularity is mirrored in a simple optics setup involving a lens with spherical aberration.\footnote{We thank Douglas Stanford's encouragement to look for an optics analogy, and Simon Caron-Huot for inspiration for this style of thinking.} This is pictured in Figure \ref{fig:lens}. The idea is to treat each incoming ray hitting the lens as a wave, and then superpose these waves to compute the resulting interference pattern on the horizontal $z$-axis. 

Consider the ray that moves along the $z$-axis, and a ray that hits the lens at a distance $b$ above the $z$-axis. Both waves experience different phase shifts as they travel inside the lens. An ideal lens is built such that both these waves have the same phase when they arrive at the focus $z=f$. Thus an ideal lens may be identified with a characteristic function $W(b)$ that satisfies
\begin{align}
\label{eq:ideal_lens}
    f=\sqrt{f^2+b^2}+W(b) \implies W(b)\sim b^2,\, b\ll f.
\end{align}
The two sides above are the optical path lengths of the two waves at $z=f$. This implies that for an ideal lens, the electric field amplitude (or the square root of the intensity of light) at a general point $z$ close to $f$ is given by
\begin{align}
    \mathcal E(z)
\propto \int_{-B}^{B}\d b\,|b|^{2p+1}
\exp\left[-\frac{ik(z-f)}{2f^2}b^2\right].
\end{align}
Here $|b|^{2p+1}$ is the incident amplitude of light, and $B$ is the range of $b$ for which the lens is illuminated.
Now, for a lens with spherical aberration, light rays don't all arrive at the focus and \eqref{eq:ideal_lens} fails. For the simplest deviation away from an ideal lens that is symmetric under $b\to -b$ and preserves the focal length, $W(b)$ receives a correction that is proportional to $b^4$. The electric field amplitude is then given by
\begin{align}
E(z)
&\propto \int_{-B}^{B}\d b\,|b|^{2p+1}
\exp\left[-\frac{ik(z-f)}{2f^2}b^2+\frac{ika}{4}b^4\right]
\nonumber\\
&= \int_0^{B^2}\d q\,q^p
\exp\left[-\frac{ik(z-f)}{2f^2}q+\frac{ika}{4}q^2\right],
\quad q=b^2,
\end{align}
with $a>0$.
This is the same integral we got for the correlator in \eqref{eq:more_gaussian}, but with $\Im c_\eta=0$ and a finite upper bound! This means that the squared magnitude of the correlator we computed maps to the intensity of light measured in this example. The corrections due to time delay of the null geodesic are analogous to the corrections due to spherical aberration, which makes the light rays hit the $z$ axis after $z=f$. The upper bound $q\sim B^2$ may also be seen as akin to the upper bound in energies due to black hole formation.

\section{A toy example}
\label{sec:toy}
\begin{figure}
    \centering
    \includegraphics[width=0.2\linewidth]{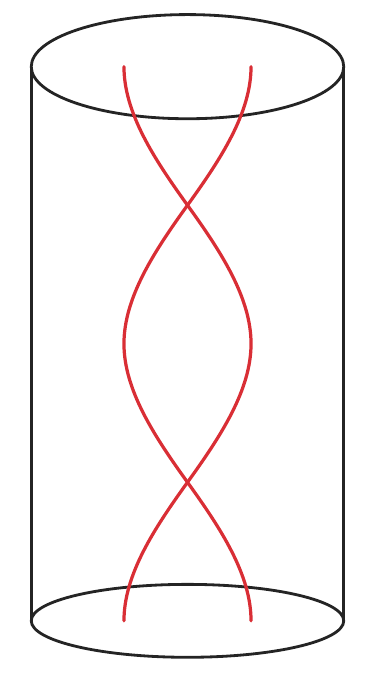}
    \caption{As a toy setup for studying the resolution of bulk cone singularities, we consider two point particles oscillating about the center of empty AdS$_3$. This is the simplest example giving rise to a bulk cone singularity in $2+1$D, where we can do a controlled backreaction computation. For instance, geodesics travelling through the background of a single point particle in AdS arrive on the boundary lightcone \cite{Berenstein:2022ico}.}
    \label{fig:cylinder_setup}
\end{figure}

As a concrete toy model for matter in the bulk we consider two particles of equal mass oscillating about the center of AdS, as shown in Figure \ref{fig:cylinder_setup}. Each particle is boosted by some amount, with the two particles boosted in opposite directions. In this setup we will show how to calculate the energy-dependent time delay of the null geodesic due to its own backreaction. As explained in Section \ref{sec:resolution}, this effect leads to the resolution of the bulk cone singularity.

\subsection{Description of the geometry}

 The geometry of these two particles is given by cutting out conical deficit angles around the particles in a specific way. Before describing this geometry, it is instructive to describe the geometry created by one particle that oscillates about the center of AdS, as derived in \cite{Lindgren_2016}. 

The idea is that we start with the geometry for a particle that is stationary at the center of AdS, and do a boost transformation. The amount of boost is parameterized by $\zeta$, and the mass of the particle by $\nu$. The backreaction of a stationary particle involves introducing a deficit angle into the empty AdS geometry, whose metric is
\begin{align}
\label{eq:globalAdS}
    \d s^2=-(r^2+1)\d t^2+\frac{\d r^2}{r^2+1}+r^2\d\phi^2.
\end{align}
The backreaction of the oscillating particle is then given by cutting out a boosted wedge from the empty AdS spacetime.
A wedge is a part of spacetime bounded by two surfaces, $v^+$ and $v^-$, which are identified. 
These surfaces extend all the way to the boundary, cutting out portions of it. The geometry looks as follows, with the right figure depicting a cross section of the cylinder on the left:
\begin{equation}
    \includegraphics[width=3.6in]{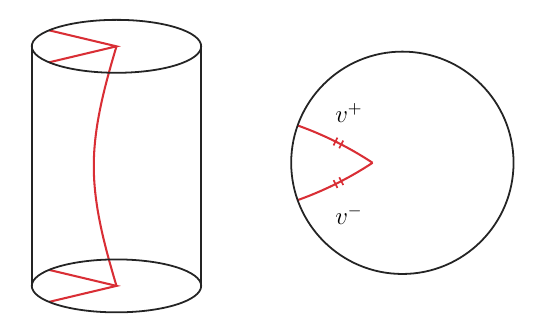}.
\end{equation}
Here we have focused on half an oscillation period.
The boundary restrictions of $v^+$ and $v^-$ are labelled $w^+$ and $w^-$, and given by
\begin{align}
    &w^-:\ \ \phi=-\Gamma+\cos^{-1}(-\sin t \sin\Gamma\tanh\zeta)\\
    &w^+:\ \ \phi=\Gamma+\cos^{-1}(\sin t \sin\Gamma\tanh\zeta).
\end{align}
In the above equations we defined $\Gamma\equiv \tan^{-1}(\tan\nu\cosh\zeta)$, which is half of the angle between $v^+$ and $v^-$ at $t=0$. We are interested in $t\in [0,\pi]$, with $\phi$ ranging from $-\pi$ to $\pi$.\footnote{Branches of inverse trigonometric functions in various places are chosen to obtain answers in this range.}
The boundary restriction of the geometry looks like
\begin{equation}
    \includegraphics[width=3.7in]{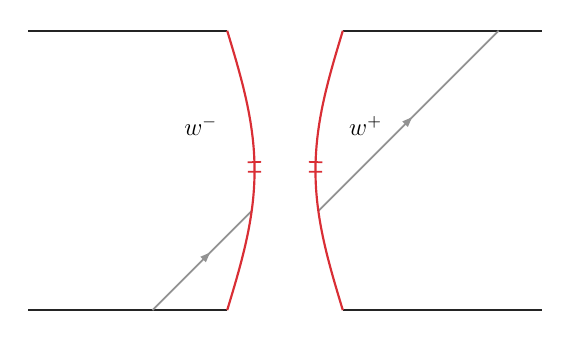}.
\end{equation}
In the above figure, the grey line is a null geodesic that moves along the boundary and hits $w^-$. Since the two wedge boundaries are identified, it then skips ahead to $w^+$ and continues.
Using the above we can build the geometry for two particles oscillating about the center of AdS, boosted in two opposite directions:
\begin{equation}
    \includegraphics[width=3.6in]{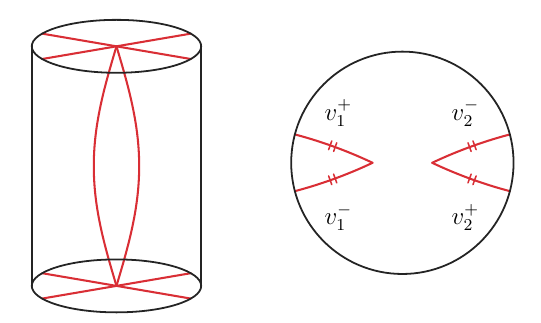}.
\end{equation}
In this case we have four curves bounding the wedges, two for each particle. Their boundary restrictions are given by
\begin{align}
    &w_1^-:\ \ \phi=-\Gamma+\cos^{-1}(-\sin t \sin\Gamma\tanh\zeta)\\
    &w_1^+:\ \ \phi=\Gamma+\cos^{-1}(\sin t \sin\Gamma\tanh\zeta)\\
    &w_2^-:\ \ \phi=-\Gamma-\cos^{-1}(\sin t \sin\Gamma\tanh\zeta)\\
    &w_2^+:\ \ \phi=\Gamma-\cos^{-1}(-\sin t \sin\Gamma\tanh\zeta).
\end{align}
As pictured on the boundary, these curves look as below:
\begin{equation}
    \includegraphics[width=3.8in]{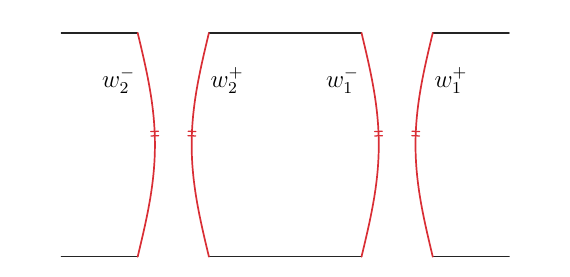}.
\end{equation}

\subsubsection*{The embedding of the boundary}
For our discussion it is important to note that in the metric presented in this gauge, the AdS boundary is not embedded at a constant value of $r$. Its embedding at some $r_*=\frac{f(t,\phi)}{\varepsilon}$ is determined by the fact that it is topologically a cylinder. The form of $r_*$ is determined by three conditions. First, the induced metric on the boundary in some intrinsic boundary coordinates $(T,\theta)$ should be a constant multiplied by $-dT^2+d\theta^2$. Second, the curves $w^\pm_{1,2}$ restricted to the boundary should be mapped to straight lines at constant $\theta$. Third, the points identified with each other in a given set of two lines should have the same $T$ coordinate. This is required so that the two segments of the boundary join smoothly at these curves, and a null geodesic doesn't bend or get delayed/advanced as it crosses them. These conditions are satisfied by the map\footnote{A similar map was used in a different context in \cite{Batra:2024qju}.} 
\begin{align}
\label{eq:map1}
    \tan{t}&=\frac{\sin(2\nu'T/\pi)\cosh{\zeta'}-\cos(2\nu'\theta/\pi-2\nu')\sinh\zeta'}{\cos(2\nu'T/\pi)}\\
\label{eq:map2}    \tan\phi&=\frac{\sin(2\nu'\theta/\pi-2\nu')}{-\sin(2\nu'T/\pi)\sinh\zeta'+\cos(2\nu'\theta/\pi-2\nu')\cosh\zeta'}
\end{align}
for the part of the cylinder bounded by $w_1^+$ and $w_2^-$, and the map
\begin{align}
\label{eq:map3}    \tan{t}&=\frac{\sin(2\nu'T/\pi)\cosh{\zeta'}-\cos(2\nu'\theta/\pi)\sinh\zeta'}{\cos(2\nu'T/\pi)}\\
 \label{eq:map4}   \tan\phi&=\frac{\sin(2\nu'\theta/\pi)}{-\sin(2\nu'T/\pi)\sinh\zeta'+\cos(2\nu'\theta/\pi)\cosh\zeta'}
\end{align}
for the part of the cylinder bounded by $w_2^+$ and $w_1^-$. The first map takes the bounding curves $w_1^+$ and $w_2^-$ to $\theta=\pi/2$ and $\theta=3\pi/2$ respectively, and the second map takes the bounding curves $w_2^+$ and $w_1^-$ to $\theta=-\pi/2$ and $\theta=\pi/2$ respectively. In the above equations we defined
\begin{align}
    \Gamma'&=\frac{\pi}{2}-\Gamma,\\
    \tanh\zeta'&=\tanh\zeta\tan\Gamma,
\end{align} 
where $\Gamma'\equiv\tan^{-1}(\tan\nu'\cosh\zeta')$.\footnote{Note that for $\zeta'$ to be real we require $\tan\nu\sinh\zeta<1$.}
The location of the cutoff boundary is then given by 
\begin{equation}
r_*(t,\varphi)
=
\frac{1}{\varepsilon}
\begin{cases}
\displaystyle
\frac{\pi}{2\nu'}
\left[
\cos^2 t
+
\left(
\cosh\zeta'\sin t-\sinh\zeta'\cos\varphi
\right)^2
\right]^{-1/2},
& (t,\varphi)\in\mathcal{B}_1,
\\[2.5mm]
\displaystyle
\frac{\pi}{2\nu'}
\left[
\cos^2 t
+
\left(
\cosh\zeta'\sin t+\sinh\zeta'\cos\varphi
\right)^2
\right]^{-1/2},
& (t,\varphi)\in\mathcal{B}_2.
\end{cases}
\end{equation}
Here $\mathcal{B}_1$ is the part of the cylinder boundary between $w_1^+$ and $w_2^-$, and $\mathcal{B}_2$ is the part of the cylinder boundary between $w_2^+$ and $w_1^-$.
To find this expression, we plug in $r=r_*$ into the global AdS metric \eqref{eq:globalAdS} and solve for the $r_*$ that gives the flat induced metric on the boundary, to leading order in $\varepsilon$.

 The way we arrived at the above map is as follows. The embedding of $\mathcal{B}_1$ is the same as the embedding of the cylinder boundary in the geometry of a single particle with parameters $(\nu',\zeta')$ boosted in the $\theta=\pi$ direction. Similarly, the embedding of $\mathcal{B}_2$ is the same as the embedding of the cylinder boundary in the geometry of a single particle with parameters $(\nu',\zeta')$ boosted in the $\theta=0$ direction. 
The remaining requirement is that these embeddings connect smoothly at the locations of the wedges, which one can check is true by noting that the map takes $w_1^+$ and $w_1^-$ to the same $\theta$, and $w_2^+$ and $w_2^-$ to the same $\theta$.

\subsection{Derivation of bulk cone singularities}
\label{sec:toy_derivation}
Our first goal is to show that there exists at least one extra singularity that doesn't live on the light cone. For this, we want to show that there exists a point $X$ such that null geodesics that leave $X$ and go through the bulk are time delayed compared to those that go through the boundary. Showing this in the geometry we described is simple using the above pictures, and is illustrated in the figure below:
\begin{equation}
    \includegraphics[width=3.8in]{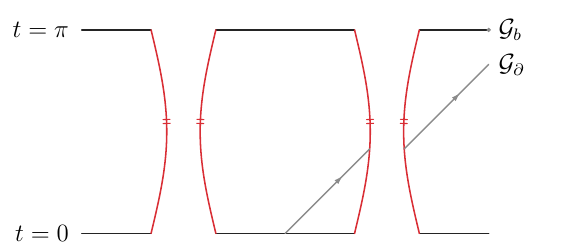}.
\end{equation}
Here, the null geodesic that goes through the bulk is $\cG_b$ and the one that goes through the boundary is $\cG_\partial$. In the above picture, the one that goes through the bulk leaves the boundary at $t=0$, and since it travels through a locally AdS spacetime, it reaches the boundary again at $t=\pi$. 

We can explicitly compute where $\cG_b$ reaches the boundary again in the intrinsic coordinates $(T,\theta)$ of the boundary cylinder. Let the geodesic start at $(t_0,\phi_0)=(0,0)$. 
The null geodesic through the bulk arrives at the boundary at $(t_B,\phi_B)=(\pi,\pi)$. The null geodesic through the boundary arrives earlier at $(t_A,\pi)$, with $t_A<t_B$. We can then find $(T_A,\theta_A)$ and $(T_B,\theta_B)$ explicitly from the maps \eqref{eq:map1}-\eqref{eq:map4}. $(T_A,\theta_A)$ equals $\left(\pi+\frac{\pi}{2\nu'}\sin^{-1}(\tanh\zeta'),\pi\right)$, since in the $(T,\theta)$ coordinates the starting point of the geodesic $(t,\phi)=(0,0)$ maps to $(T,\theta)=\left(\frac{\pi}{2\nu'}\sin^{-1}(\tanh\zeta'),0\right)$. For the geodesic through the bulk, we have that $\theta_B$ equals $\pi$, whereas
\begin{align}
\label{eq:TB}
    T_B=\frac{\pi}{2\nu'}\left[\pi-\sin^{-1}(\tanh\zeta')\right],
\end{align}
so that the time delay equals
\begin{align}
\label{eq:toy_t_delay}
    \Delta T=\frac{\pi}{2\nu'}\left[\pi-2\sin^{-1}(\tanh\zeta')\right]-\pi.
\end{align}
Using some trigonometric manipulations one can show that this is indeed positive. Indeed, the parameter relations imply
$\sin\nu'=\cos\nu/\cosh\zeta'<1/\cosh\zeta'$,
and hence $\nu'<\pi/2-\sin^{-1}(\tanh\zeta')$,
which makes \eqref{eq:toy_t_delay} strictly positive.

\subsection{Correction due to the time delay of the null geodesic}
The basic effect in the presence of the null geodesic is a further change in the embedding of the boundary. Before deriving this change in embedding, we need to understand the backreaction of the null geodesic on the spacetime. Similar to the massive particles, the null geodesic cuts out a wedge from the AdS spacetime. At different moments in time, this may be pictured as:
\begin{equation}
    \includegraphics[width=5.2in]{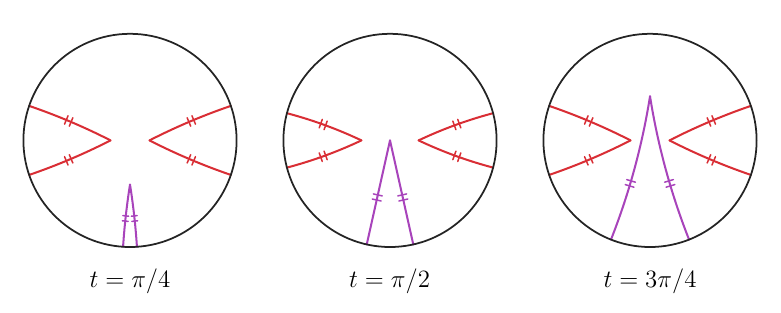}.
\end{equation}
The boundary restrictions of the curves $n^+$ and $n^-$ are given by
\begin{align}
    &n_+:\ \ \phi=\epsilon-\sin^{-1}(\cos t \sin\epsilon)\\
    &n_-:\ \ \phi=-\epsilon+\sin^{-1}(\cos t \sin\epsilon),
\end{align}
so that the picture of the boundary now becomes
\begin{equation}
    \includegraphics[width=4in]{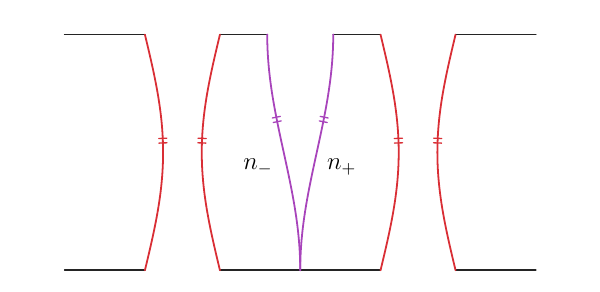}.
\end{equation}
Here $\epsilon$ ranges from $0$ to $\pi/2$, with $\tan\epsilon$ representing the dimensionless energy of the point particle.\footnote{The physical energy $E$ may be found by restoring $G_N$ and is instead given by $\tan\epsilon\sim G_N E$.}
It is difficult to compute the exact change in the embedding of the boundary due to the backreaction of the null geodesic, and hence the correction to the map from $(t,\phi)$ to $(T,\theta)$ coordinates. However, we can still derive the change in the time $T_B$ at which the null geodesic $\mathcal{G}_b$ that goes through the bulk reaches the boundary again as follows.\footnote{We thank Douglas Stanford for suggesting this method.}

First note that right at $t=0$ when the null geodesic is sent into the bulk, the map from $(t,\phi)$ to $(T,\theta)$ is the same as the one outlined in the previous subsection. This is needed for the metric on the boundary to be continuous across the $t=0$ slice. 
Now, we can imagine sending null geodesics from the point $(t_B,\phi_B)$ back into the past. These intersect the line at $t=0$ at the angular locations $\phi_\epsilon^+$ and $\phi_\epsilon^-$ that depend on $\epsilon$. This is visualized below:
\begin{equation}
\label{eq:backwards}
    \includegraphics[width=4.8in]{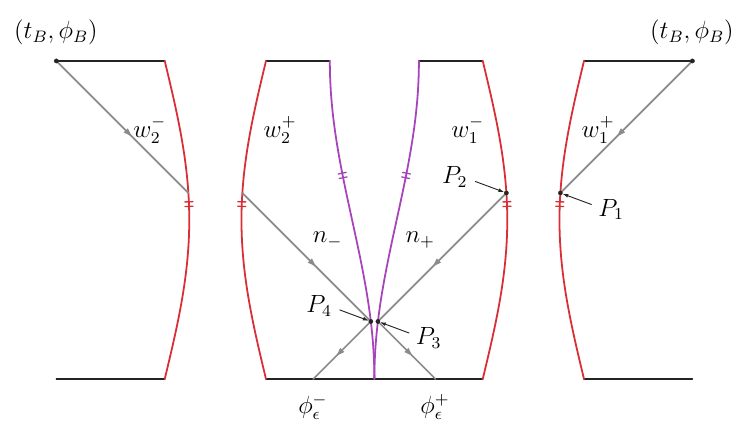}.
\end{equation}
Since we know the map from $(t,\phi)$ to $(T,\theta)$ on the $t=0$ slice, we can translate the points $P_\epsilon^-=(0,\phi_\epsilon^-)$ and $P_\epsilon^+=(0,\phi_\epsilon^+)$ to the $(T,\theta)$ coordinates. We can then switch to the point of view of the $(T,\theta)$ coordinates, and send null geodesics from $P_\epsilon^-$ and $P_\epsilon^+$ into the future. The point at which these null geodesics intersect for the second time is then $(T_B,\theta_B)$ with the backreaction of the null geodesic taken into account.

To carry this out, we first compute the point $P_1$ shown in \eqref{eq:backwards} by finding the intersection of the null ray sent into the past from $(t_B,\phi_B)$ with the curve $w_1^+$. This null ray satisfies $\phi=t$, which determines the point of intersection to be
\begin{align}
    P_1: t_1=\phi_1=\tan^{-1}\left[\frac{\cot\Gamma}{\tanh\zeta-1}\right].
\end{align}
$P_2$ is then given by
\begin{align}
    P_2: t_2=\phi_1,\ \phi_2=\pi-\phi_1.
\end{align}
To find $P_3$, we find the intersection of the null ray from $P_2$ to $P_3$ (given by $\phi=t+\pi-2\phi_1$) with $n_+$. This gives
\begin{align}
    P_3: t_3=\tan^{-1}\left[\frac{\sin(\epsilon+2\phi_1-\pi)-\sin\epsilon}{\cos(\epsilon+2\phi_1-\pi)}\right],\ \phi_3=t_3+\pi-2\phi_1.
\end{align}
This implies $P_4=(t_3,-\phi_3)$. The null ray starting at $P_4$ then meets the $t=0$ slice at 
\begin{align}
    \phi_{\epsilon}^{-}=2\phi_1-\pi-2t_3.
\end{align}
At this slice, using the map from $(t,\phi)$ to $(T,\theta)$ in the absence of the null geodesic is valid. Using \eqref{eq:map3} and \eqref{eq:map4}, the points with $t=0$ and $\phi=\phi_\epsilon^\pm$ map to
\begin{align}
    P_\epsilon^-&: T_\epsilon^-=\frac{\pi}{2\nu'}\sin^{-1}\left[\tanh\zeta'\cos\!\left(\frac{2\nu'}{\pi}\theta_\epsilon^-\right)\right],\ \theta_\epsilon^-=\frac{\pi}{2\nu'}\tan^{-1}\left(\frac{\tan\phi_\epsilon^-}{\cosh\zeta'}\right).\\
    P_\epsilon^+&: T_\epsilon^-=T_\epsilon^+,\ \theta_\epsilon^+=-\theta_\epsilon^-.
\end{align}
A straightforward calculation on the intrinsic boundary cylinder then gives that the null geodesics emitted towards each other from the above two points intersect for a second time at
\begin{align}
    T_B^\epsilon=T_\epsilon^-+\pi-\theta_\epsilon^-,\ \theta_B^\epsilon=\pi.
\end{align}
This is pictured below:
\begin{equation*}
    \includegraphics[width=3.4in]{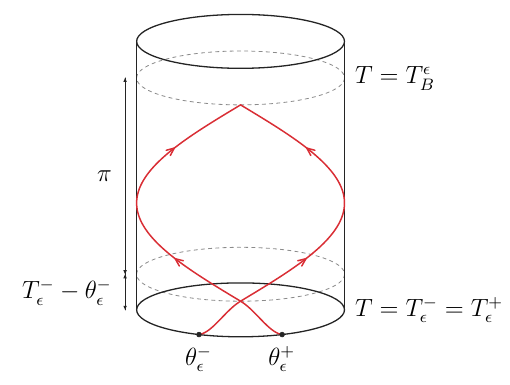}
\end{equation*}
Thus compared to the time $T_B$ in \eqref{eq:TB}, taking the backreaction of the null geodesic into account gives an additional time delay 
\begin{align}
\Delta T_B
\equiv T_B^\epsilon-T_B \nonumber=
T_\epsilon^-+\pi-\theta_\epsilon^-
-\frac{\pi}{2\nu'}
\left[
\pi-\arcsin\!\left(\tanh\zeta'\right)
\right].
\end{align}
The important point is that $\Delta T_B$ as computed above depends on $\epsilon$ and so on the energy of the null geodesic. We can see it is positive by expanding it for small $\epsilon$:
\begin{equation}
\Delta T_B
=
\frac{2\pi\cos^2\phi_1}
{\nu'\!\left[
\cosh\zeta'
-\sinh\zeta'\sin(2\phi_1)
\right]}
\,\epsilon
+O(\epsilon^2).
\end{equation}
Combined with the analysis in Section \ref{sec:resolution}, this shows that the bulk cone singularity predicted in this model in Section \ref{sec:toy_derivation} gets resolved due to the backreaction of this null geodesic. In the Section \ref{sec:resolution} discussion we fixed the endpoint $x$ of the null geodesic, and chose the starting point of the geodesic so that it has initial energy $E$ and arrives at the point $x$. This energy-dependent time delay says that this starting point gets shifted earlier by an energy-dependent amount.

Note that the treatment of Section \ref{sec:resolution} is in a gauge where the metric is asymptotically AdS and changes as we move into the bulk. In this section, we computed the additional time delay of the null geodesic by working in this simpler gauge where the metric is always locally AdS --- apart from the additional wedge created by the null geodesic, the boundary embedding is the only degree of freedom affected by backreaction. The fact that the additional time delay is a gauge-invariant quantity allows us to use it in the computations outlined in Section \ref{sec:resolution}.

\subsubsection*{Black hole formation and inelastic effects}
An interesting property of this toy model is that we can explicitly see black hole formation occur. Consider progressively increasing the energy $\epsilon$ of the null particle. At some critical energy $\epsilon_*$, the boundary restrictions of the curves $w_{1,2}^\pm$ and $n^\pm$ look as follows:
\begin{equation}
    \includegraphics[width=0.68\linewidth]{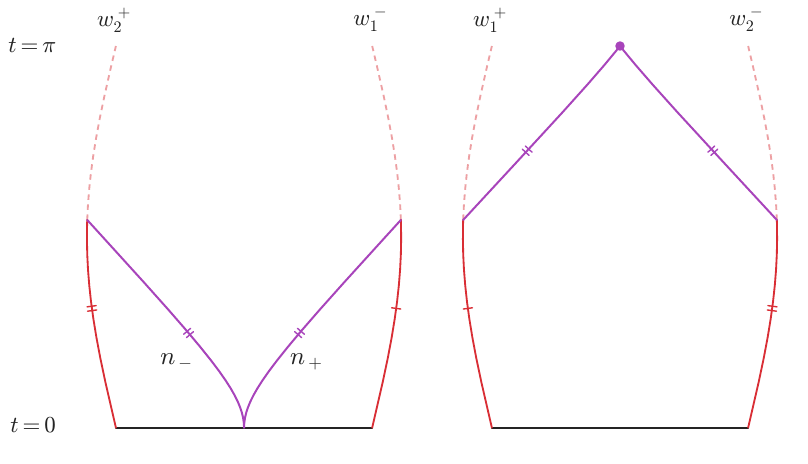}.
\end{equation}
The point at $t=\pi$ where the curves $n_+$ and $n_-$ meet again lies at $T=\infty$ in the intrinsic boundary coordinates. 
This may be seen by sending two null rays backwards from a point approaching this point, and noting that the number of times they intersect each other before reaching the $t=0$ slice approaches infinity. Thus past this critical energy $\epsilon_*$, the null geodesic sent through the bulk does not reach the boundary again, and the bulk is in a regime where a black hole has formed. Such black hole formation is discussed in \cite{Lindgren_2016} from the bulk point of view.

Let us also comment on how other inelastic effects appear in this model. As explained in Section \ref{sec:inelastic_effects}, these give the coefficient $c_\eta$ a positive imaginary part. We expect this imaginary part to be non-zero in this setup, for the following reason. The backreaction of the null geodesic leads to a different embedding of the boundary in the future, compared to the past. Taking the overlap of this state in the future with the boundary graviton vacuum should then lead to a prefactor with magnitude less than one, that decreases as the energy of the null geodesic increases. In general this would give the $\exp(-\# G_N E^2)$ correction that equips $c_\eta$ with a positive imaginary part. It would be nice to show its existence in a more concrete calculation.

\section{Discussion}
\label{sec:disc}
In this paper we have argued that at non-zero $G_N$, the bulk cone singularity of AdS/CFT correlators at fixed spatial momentum gets resolved. We showed that the dominant physical effect causing this resolution is the gravitational backreaction of the null geodesic causing the singularity. Since this null geodesic is not prompt, parts of it in the past can influence parts of it in the future, which causes an energy-dependent correction to its trajectory. Additionally, imposing the absence of gravitational radiation in the final state adds an inelastic contribution. These corrections fully resolve the singularity. 

We studied the form of the correlator in a region of size $\mathcal{O}(\sqrt{G_N})$ around the resolved singularity. It would be interesting to understand what happens globally in the complex $t$ plane ---  does one obtain a structure of Stokes lines that mirrors the structure identified in \cite{Batra:2026gbq} for the bouncing geodesic singularity at non-zero $G_N$? Locally the resolution of the singularity is given by a Gaussian smoothing, which matches the local structure of Stokes lines in \cite{Batra:2026gbq}. Similar to the bouncing geodesic case, does one of these Stokes lines cause a Stokes phenomenon near the point $t=0$? Does the function get exponentially large in $1/G_N$ in some direction around the resolved singularity in the complex $t$ plane?

 In the setup of \cite{Batra:2026gbq}, the saddle point that gets picked up across a Stokes line near $t=0$, as well as the region where the function is exponentially large, encode UV information about the geometry. This makes these structures especially interesting --- in this case, do they teach us something about black hole formation? Similar to the bouncing geodesic, does accessing this UV information require us to analytically continue to negative Euclidean time? We saw a hint of this in Section \ref{sec:corrected_correlator}, where the corrected correlator grows in the direction of negative Euclidean time. Making progress on the above questions requires (and may help with) a better understanding of the inelastic effects in this setup. Perhaps a good avenue to identify the above structures could be to study the resolution of these singularities in CFT$_2$.

 In general it is mysterious how bulk cone singularities arise in the boundary theory. It would be interesting to shed light on the boundary physics that gives rise to them. Do these occur only in holographic theories, or are they a general feature of chaotic systems? A first step is to reproduce them in a strongly coupled quantum system, perhaps along the lines of \cite{Dodelson:2024atp,Dodelson:2025jff,Dodelson:2026gak}.

 The most interesting question is the fate of the backreacted bouncing geodesic in higher dimensions. Our analysis in Section \ref{sec:resolution} does not apply to this geodesic since it probes a region of high curvature. A related setup that combines such singularities with the bouncing geodesic singularity is the case of the charged black hole considered in \cite{Ceplak:2025dds,AliAhmad:2026wem}. The null geodesic causing a singularity in this case bounces off a timelike singularity, and is not prompt. Does the mechanism that resolves the singularity in this paper resolve the bouncing geodesic singularity in that case as well?

\subsection*{Acknowledgments}
I am very grateful to Jan Boruch, Simon Caron-Huot, Joydeep Chakravarty, Yiming Chen, Matthew Dodelson, Mason Kamb, Adam Levine, Raghu Mahajan, Steve Shenker and Nico Valdes-Meller for helpful discussions, and to Steve Shenker and Douglas Stanford for comments on the draft. This work benefitted from a ChatGPT Pro subscription, and used AI for help with technical details. I am especially thankful to Douglas Stanford for extended discussions, advice and suggesting many important ideas, not all of which could be summarized in the footnotes. In particular, Douglas asked me the following question in 2023 that started this project: can the backreaction of a null geodesic resolve the bulk cone singularity? GB was supported in part by NSF Grant PHY-2310429. 

\appendix

\section{More on the van Vleck-Morette determinant}
\label{sec:van_vleck}
A good reference for understanding the van Vleck-Morette determinant introduced in Section \ref{sec:energy_integral} is \cite{Caron-Huot:2025hmk}. In this section, we will outline the information about it that is needed for the argument in this paper. When only one endpoint of the geodesic under consideration is fixed, this determinant is given by
\begin{align}
\label{eq:vVM_det}
\mathcal D(p;x)
=
\frac{\sqrt{-g(X)}}{\sqrt{-g(x)}}
\det\nolimits^{\prime}\!\left(
\frac{\partial^2 S_\eta(p;x)}
{\partial p_\mu\,\partial x^M}
\right).
\end{align}
Here $x^M$ is one endpoint of the geodesic, $p^\mu$ is the momentum of the other endpoint at the boundary, and $X$ is the location of that endpoint. $g(X)$ is the determinant of the conformal boundary metric, and $g(x)$ is the determinant of the bulk metric.
Thus $\mathcal{D}$ is a measure of how the action of the geodesic changes locally upon changing the endpoint of one end of the geodesic, or the momentum of the other end. The determinant above is that of a rectangular matrix. The definition of this determinant is decided by
\begin{align}
\label{eq:D_def}
    \frac{d^{d+1}P\,\delta(P^2+m^2)}{\sqrt{-g(x)}}
\equiv
\mathcal D(p;x)\,\frac{d^d p}{\sqrt{-g(X)}}.
\end{align}
Here $P$ is the full bulk momentum of the geodesic.
Using this definition we can show the scaling $\sqrt{\mathcal D_\eta}\propto E^{-1/2}$ claimed in Section \ref{sec:singularity_der}. Consider the rescaling $p_\mu\to a p_\mu$ with $a>0$. For a null geodesic, this also rescales $P_M\to a P_M$.  Under this rescaling, the measures scale as
\begin{align}
    d^{d+1}P\delta(P^2)\to a^{d-1} d^{d+1}P\delta(P^2),\quad d^dp\to a^d d^dp.
\end{align}
Here we ignored the $m^2$ inside the $\delta$ function since we are considering large $P$. Comparing the two sides of \eqref{eq:D_def} then implies
\begin{align}
    \mathcal{D}(ap;x)=a^{-1}\mathcal{D}(p;x).
\end{align}
So for a geodesic with $p^\mu=(E,\vec 0)$, we have that $\mathcal{D}\propto E^{-1}$, as required. 

Using the expression in \eqref{eq:vVM_det} we can also see how $\mathcal{D}$ changes upon a change in the metric $g_{\mu\nu}=g_{\mu\nu}^0+h_{\mu\nu}$. As argued above, $\mathcal{D}$ equals $1/E$ times a function of the metric $g$. In Section \ref{sec:backreaction} we argued why to leading order in $G_N$, $h\sim G_N E$. This implies that $\mathcal D$ has the following Taylor series expansion in $G_N E$:
\begin{align}
    \mathcal D\sim \mathcal D_0(1+\# G_N E + \# G_N^2 E^2+\cdots),
\end{align}
with $\mathcal D_0\sim 1/E$.

\section{The coupling of the null geodesic to metric fluctuations}
\label{sec:diagrams}
The action governing the interactions of the metric and the null geodesic (given by a massless particle) is given by
\begin{equation}
S_{\rm pp}
=
\frac12\int d\lambda
\left[
g_{\mu\nu}(X)
\dot X^\mu\dot X^\nu
\right].
\end{equation}
Here we set the einbein field $\eta=1$, which gauge fixes the reparametrization invariance of the worldline. To find how various vertices scale with the momenta, it is better to work with the equivalent action
\begin{align}
\label{eq:app_pp_action}
    S_{\rm pp}[X,P,\eta,g]
    = \int \d\lambda \left[
        P_\mu \dot X^\mu
        - \frac{\eta}{2} g^{\mu\nu}(X) P_\mu P_\nu
    \right].
\end{align}
We can expand the metric around the background solution without the null geodesic as
\begin{align}
    g_{\mu\nu}=\bar g_{\mu\nu}+\sqrt{G_N}h_{\mu\nu}.
\end{align}
The factor of $\sqrt{G_N}$ ensures that the propagators of the $h_{\mu\nu}$ field are $\mathcal{O}(1)$.
Plugging this into \eqref{eq:app_pp_action}, and using  
\begin{equation}
g^{\mu\nu}
=
\bar g^{\mu\nu}
-\sqrt{G_N}h^{\mu\nu}
+G_N h^\mu{}_\rho h^{\rho\nu}
-G_N^{3/2}h^\mu{}_\rho h^\rho{}_\sigma h^{\sigma\nu}
+\cdots,
\end{equation}
we find that the vertex coupling $n$ gravitons to the point particle is schematically
\begin{align}
    S_{\rm pp}^{(n)}
    \sim G_N^{n/2} \int \d\lambda\,
    P_\mu (h^n)^{\mu\nu} P_\nu.
\end{align}
Now, it is helpful to separate the metric perturbations into the trace and traceless perturbations. The main result of this section is that a traceless perturbation couples to the point particle as
\begin{align}
    G_N^{n/2} \int \d\lambda\,
    P_\mu (h_{\rm TL}^{\,n})^{\mu\nu} P_\nu
    \sim G_N^{n/2} E.
\end{align}
To find the scaling on the right hand side above, we used the scaling $d\lambda\sim E^{-1}$ and $P_\nu \sim E$ derived in Section \ref{sec:backreaction}.\footnote{For simplicity we are again considering an $L=0$ null geodesic.} An equivalent way to see this scaling is to write this action in terms of the coordinate time along the null geodesic, using $\d t/d\lambda \sim E$ and the fact that the background null geodesic trajectory is independent of $E$. 

Using the mass shell constraint $\bar g_{\mu\nu}\dot X^\mu\dot X^\nu=0$, the coupling of the pure trace perturbation $h_{\rm tr}^{\mu\nu}
\sim\bar g^{\mu\nu}h$ to the massless particle instead evaluates to zero. A similar argument shows that vertices involving both pure trace and traceless perturbations have the same scaling $\sim G_N^{n/2} E$. 

Note that we did not consider fluctuations of the point particle in our analysis since those are suppressed in the region where the WKB approximation is valid (except Gaussian fluctuations, which are included through the van Vleck–Morette prefactor). In the region where it breaks down, we instead match to the exact solutions of the wave equation, as explained in Section \ref{sec:energy_integral}.

\section{Gravitational radiation from a null geodesic in AdS$_5$-Schwarzschild}
\label{sec:radiate}
As an example we consider a null geodesic in the AdS$_5$ Schwarzschild background that starts at the boundary, goes to a minimum radius $r_{\rm{min}}$, and then returns to the boundary.\footnote{We thank ChatGPT for this computation.} It has angular momentum $L$ and energy $E$, with $b\equiv L/E$. The metric is given by
\begin{equation}
    ds^2 = -\left(1+r^2-\frac{\mu}{r^2}\right)dt^2
    + \frac{dr^2}{1+r^2-\frac{\mu}{r^2}}
    + r^2 d\Omega_3^2,
\end{equation}
with $\mu$ being related to the black hole mass. We also define the quantities
\begin{align}
    f(r) = 1+r^2-\frac{\mu}{r^2},
    \quad
    F(r)=1-\frac{b^2f(r)}{r^2},\quad 
    t(r) = \int_{r_{\min}}^r
    \frac{dr'}{f(r')\sqrt{F(r')}} .
\end{align}
We can write the perturbed metric as $g=g^{(0)}+\sqrt{32\pi G_N}\,h$, and focus on the
mode $h_{ij}=r^2R_\omega(r)e^{-i\omega t}\mathbb T_{ij}$. Here $\mathbb T_{ij}$ is a unit-normalized, transverse traceless
$\ell=2$ harmonic on $S^3$, chosen so that
$\mathbb T_{\phi\phi}=1/(\pi\sqrt{3})$ along the ray. The function $R_\omega(r)$ obeys Eq. (5.3) of \cite{Kodama:2003jz}:
\begin{equation}
\label{eq:KI}
    \frac{d}{dr}\left(r^3 f(r)\frac{dR_\omega}{dr}\right)
    + r^3\left(\frac{\omega^2}{f(r)}-\frac{8}{r^2}\right)
    R_\omega(r)=0.
\end{equation}
We choose the solution that behaves as $R_\omega\sim r^{-4}$ at infinity, since the other solution changes the boundary metric. Since our interest is in showing a non-zero probability of gravitational radiation, the normalization of the solution does not matter.

The formula for the amplitude of graviton emission in Eq. (62) of \cite{Bernar:2017kug}
applied to the full null geodesic then gives
\begin{align}
\mathcal A_\omega
    &= \frac{i\sqrt{32\pi G_N}}{2}
    \int d^5x\,\sqrt{-g}\,
    h_{\mu\nu,\omega}^{*}(x)\,T^{\mu\nu}(x)
    \nonumber\\
&=\frac{i\sqrt{32\pi G_N}}{2}
  \int \mathrm{d}\lambda\,p^\mu p^\nu h_{\mu\nu}^*
\nonumber\\
&=\frac{i\sqrt{32\pi G_N}\,Eb^2}{\pi\sqrt{3}}
  \int_{r_{\min}}^\infty
  \frac{\mathrm{d}r\,R_\omega^*(r)\cos[\omega t(r)]}
       {r^2\sqrt{F(r)}}.
\end{align}
Using the solution to \eqref{eq:KI}, one can check that this is non-zero, and proportional to $\sqrt{G_N} E$ at fixed $b$.

\bibliography{main}
\bibliographystyle{JHEP}
\end{document}